\documentclass[12pt]{article}
\usepackage{amsfonts,latexsym}
\begin{document}
\newcommand{\de}{\delta}\newcommand{\ga}{\gamma}
\newcommand{\e}{\epsilon} \newcommand{\ot}{\otimes}
\newcommand{\be}{\begin{equation}} \newcommand{\ee}{\end{equation}}
\newcommand{\ba}{\begin{eqnarray}} \newcommand{\ea}{\end{eqnarray}}
\newcommand{\tmod}{{\cal T}}\newcommand{\amod}{{\cal A}}
\newcommand{\bemod}{{\cal B}}\newcommand{\cmod}{{\cal C}}
\newcommand{\dmod}{{\cal D}}\newcommand{\hmod}{{\cal H}}
\newcommand{\s}{\scriptstyle}\newcommand{\tr}{{\rm tr}}
\newcommand{\einsop}{{\bf 1}}
\def\oR{R^*} \def\upa{\uparrow}
\def\R{\overline{R}} \def\doa{\downarrow}
\def\oL{\overline{\Lambda}} 
\def\nn{\nonumber} \def\dag{\dagger}
\def\be{\begin{equation}}
\def\ee{\end{equation}} 
\def\bea{\begin{eqnarray}} 
\def\eea{\end{eqnarray}} 
\def\ve{\epsilon}
\def\si{\sigma}
\def\th{\theta} \def\ga{\gamma}
\def\l{\left} 
\def\r{\right}
\def\a{\alpha} 
\def\b{\beta} 
\def\g{\gamma} 
\def\La{\Lambda} 
\def\w{\overline{w}} 
\def\u{\overline{u}}
\def\o{\overline}
\def\rr{\mathcal{R}} 
\def\T{\mathcal{T}} 
\def\L{\mathcal{L}} 
\def\M{\mathcal{M}}
\def\N{\mathcal{N}} 
\newcommand{\reff}[1]{eq.~(\ref{#1})}

\centerline{\bf{TRANSFER MATRIX EIGENVALUES OF THE }} 
\centerline{\bf{ANISOTROPIC MULTIPARAMETRIC U MODEL }} 
~~~\\
\begin{center}
{\large Jon Links$^{1}$ and Angela Foerster$^{2}$}
\vspace{1cm}~~\\
~~ \\
{\em $^1$Centre for Mathematical Physics \\
Department of Mathematics \\
The University of Queensland,  4072 \\  Australia \\
e-mail jrl@maths.uq.edu.au}
~~\\
~~\\
{\em $^2$Instituto de F\'{\i}sica da UFRGS \\
Avenida Bento Gon\c{c}alves 9500    \\
Porto Alegre, RS - Brazil \\
e-mail angela@if.ufrgs.br}~~\\
\end{center} 
\begin{abstract}
A multiparametric extension of the anisotropic $U$ model is discussed
which maintains integrability. The $R$-matrix solving the
Yang-Baxter equation is obtained through a  twisting
construction  applied to the underlying $U_q(sl(2|1))$ superalgebraic
structure which introduces the additional free parameters that arise
in the model. Three forms of Bethe ansatz solution for the transfer
 matrix eigenvalues are given which we show to be equivalent.

\end{abstract}
\vspace{1cm}
\begin{flushleft}  
\end{flushleft}

\vfil\eject
\centerline{{\bf 1. Introduction}}
~~\\

The quantum algebras, including the $\mathbb Z_2$-graded analogues 
known as quantum superalgebras, play a central role in the construction
of solutions of the Yang-Baxter equation which in turn may be used to 
construct integrable one-dimensional quantum models. Quantum algebras 
as defined originally by Jimbo and Drinfeld \cite {j,d1}
 arise as one-parameter deformations of the familiar Lie algebras in
 such a way that the resulting algebraic structure is that of a
 quasi-triangular Hopf algebra. 
The supersymmetric
generalizations  are
defined in \cite{bgz,manfred,y} and the quasi-triangularity of these Hopf
(super)algebras is discussed in \cite{kt}.
The importance of this class of algebras
 is the existence of a universal element, known as the $R$-matrix, which
 gives a solution for the Yang-Baxter equation. 
For each given solution of 
the Yang-Baxter equation, there is a well known procedure called the 
Quantum Inverse Scattering Method (QISM) \cite{fst} by which an integrable
one dimensional quantum system is obtained. One of the key steps in 
the QISM is the construction of the transfer matrix which yields 
a family of mutually commuting operators (including the Hamiltonian)
which are taken as 
constants of the motion for the system. Diagonalization of the 
transfer matrix is the main objective in the solution of the model and from 
this result many properties such as the ground state structure and elementary
excitations can be deduced. 

Through the construction of 
twisting by  elements satisfying the two co-cycle condition it became
apparent that new quasi-triangular Hopf algebras could be manufactured 
from existing ones \cite{d2}, 
and if the twisting procedure introduced additional free 
parameters then these would appear in the Hopf algebra structure
\cite{re}. Viewed 
in another perspective, there exists classes of quasi-triangular Hopf 
algebras in which each member of the class is related to all others via 
twisting. In terms of representation theory, this means that representations
of all quasi-triangular Hopf algebras in a given class are equivalent. 

On the other hand, in applications to the area of constructing integrable 
quantum chain models some subtleties emerge. In a recent work, it was shown
that in the case of Reshetikhin twists \cite{re}, 
which will be defined precisely later,
the periodic multiparametric chains can be mapped to the standard ones 
with the inclusion of a generalized twisted  boundary condition \cite{flr}. 
In that work, two physical 
examples of the anisotropic $t-J$ and $U$ models of correlated electrons
(both derived through
representations of the quantum superalgebra $U_q(sl(2|1))$) 
were constructed and also the Bethe ansatz equations obtained. From 
these solutions, it is evident that the additional parameters introduced
by twisting cannot be transformed away and thus do impact 
on the physics that these models describe. 

Subsequent to the work of Reshetikhin \cite{re}, Engeldinger and Kempf
\cite{ek} gave a more general prescription for the twisting element (or
twistor) by relaxing the triangularity property. This latter
construction opens the possibility for even more parameters to be
incorporated into the model in an integrable fashion. When applied to
the fundamental representation of the $U_q(gl(m|n))$ series, the method
of Engeldinger and Kempf reproduces the construction of Reshetikhin. For
this reason, we cannot use these more general techniques to obtain a
more general extension of the anisotropic $t-J$ model than that already
obtained in \cite{flr}. However, when applied to ``higher spin''
representations, such as the case to be considered here, the
differences between the constructions begin to emerge. 
We will show that  these additional parameters  
describe local basis
transformations for the local Hamiltonians.
This result will be illustrated for the anisotropic $U$ model
which will be discussed in some detail.
Being a fermionic model, it is necessary to derive the model in a
supersymmetric formulation of the QISM. 

Unlike the case of integrable models based on non-graded 
algebras, there exist many examples of models with an underlying 
superalgebraic structure which admit more than one Bethe ansatz solution 
\cite{fv,fk,eks,pf96,pf97,lf99,flt}. It is generally accepted that the 
non-uniqueness of solution stems from the fact that the definition
for a Lie superalgebra in terms of a system of simple roots is not 
unique \cite{fss}. For the case of the supersymmetric $t-J$ model two forms
of solution were known long ago from the works of Lai \cite{lai} and 
Sutherland \cite{s}. With the advent of the algebraic form of the 
Bethe ansatz a third form was discovered in \cite{fv,fk}. Moreover, 
Essler and Korepin showed further in \cite{fv} through an analytical argument 
that the three forms of solution for the supersymmetric $t-J$ 
were equivalent. 

For the isotropic $U$ model, as a result of sharing 
 the same supersymmetry algebra (viz.
$sl(2|1)$) as the supersymmetric $t-J$ model, there are also 
three forms of the Bethe 
ansatz solution. For the transfer matrix eigenvalues, two forms were obtained 
by Pfannm\"uller and Frahm \cite{pf96} using standard methods. The third 
form was eventually discovered also by Pfannm\"uller and Frahm \cite{pf97}
but for this case one has to resort to more sophisticated Bethe ansatz 
techniques which were developed by Abad and R\'{\i}os \cite{ar}.   

Recall that the
anisotropic $U$ model was first introduced and solved via the
co-ordinate Bethe ansatz in \cite{bkz} as a
generalization
of both the Bariev model \cite{ba} and the (isotropic) $U$ model given
in 
\cite{bglz}. It was subsequently shown that the model is also 
obtained from an $R$-matrix solution of the
Yang-Baxter equation obtained through the one-parameter family
of four dimensional (minimal typical) representations of the quantum
superalgebra $U_q(sl(2|1))$ \cite{ghlz} (see also \cite{bdgz,dglz,m,a}). 
By applying the twisting
construction to this solution of the Yang-Baxter equation we will give
the explicit form for the resulting $R$-matrix and in turn derive 
a multiparametric generalization of the
anisotropic $U$ model.

One of the main objectives of this paper is to 
present all three forms of the Bethe ansatz solution for 
the multiparametric anisotropic $U$ model including the Bethe ansatz 
equations and explicit eigenvalue expressions for the transfer matrix. 
Since the model is not based
on the fundamental representation of its supersymmetry algebra, we will 
adopt a generalization of the procedure used in the case of 
the solution of the 
integrable spin 1 chain (Fateev-Zamolodchikov model) \cite{b,bt,lfk}. In the 
model considered here this involves determining the $L$-operator
which acts on the mixed tensor space of the fundamental module with the 
four dimensional module which describe the local quantum states. A
similar
 procedure has recently been undertaken by Gruneberg \cite{g99,g00} for 
classes of $U_q(sl(2|1))$ models        
which include the usual anisotropic $U$ model. The necessary step for the 
multiparametric case is the determination of the appropriate multiparametric
$L$-operator. In principle, it may be 
possible to work directly with the $R$-matrix
for the Bethe ansatz solution in an analogous way to that used by Ramos 
and Martins in the isotropic case \cite{rm} or even the analytic Bethe
ansatz approach developed by Tsuboi for quantum superalgebras \cite{t}.
However, we will not consider these options in this work.

Once the three forms of the Bethe ansatz solution have been obtained, we will
proceed to argue that the transfer matrix eigenvalues are equivalent. 
Our approach is somewhat different than that used by  Essler and Korepin 
 for the $t-J$ model 
 \cite{fv}. We use the result that the Bethe ansatz solution for a 
$U_q(sl(1|1))$ (free fermion) model can be approached in two different
ways and then proceed to show that the three solutions for the $U$ model
have their origin in the differing forms of 
these $U_q(sl(1|1))$ eigenvalue expressions.

One of the sequent results of the Bethe analysis for the transfer matrix 
eigenvalues is that the eigenvalues of the {\it quantum} transfer matrix
are obtained with little additional effort. The quantum transfer matrix 
approach, which has primarily been developed by Kl\"umper and collaborators
\cite{k,jks,kwz}, has proved 
to be a powerful method to determine thermodynamic properties of an 
integrable model at finite temperature. In the final section of this work 
we will present the eigenvalues of the quantum transfer matrix for the 
anisotropic multiparametric $U$ model.
 
~~\\  
\centerline{{\bf 2. Quantum Inverse Scattering Method }}
~\\

We begin by reviewing the fundamental features of the QISM.
Let $R(u)\in \mathrm{End}\,V\ot V$ be a solution of the Yang-Baxter
equation
\be 
R_{12}(u-v)R_{13}(u)R_{23}(v)=R_{23}(v)R_{13}(u)R_{12}(u-v). \label{yb}
\ee 
For full generality, we consider the cases also when 
 $V$ denotes  a ${\mathbb Z}_2$-graded vector space. In such instances
it is necessary to impose the following rule for the tensor product
multiplication of matrices:
\be (a\ot b)(c\ot d)=(-1)^{[b][c]}ac\ot bd \label{rule} \ee  
for matrices $a,\,b,\,c,\,d$ of homogeneous degree. The symbol $[a]\in
{\mathbb Z}_2$ denotes the degree of the matrix $a$. 
The monodromy matrix is defined
$$\T(u)=R_{0L}(u)R_{0(L-1)}(u)....R_{01}(u) $$
from which the transfer matrix is given by 
\be t(u)=\mathrm{str}_0 \T(u). \label{tm} \ee  
A consequence of (\ref{yb}) is
$$[t(u),\,t(v)]=0,~~~\forall\,u,v\in {\mathbb C}.$$
Above $\mathrm{str_0}$ denotes the supertrace taken over the auxiliary
space which is labeled by 0.

In the usual manner the Hamiltonian associated with the transfer matrix
is defined by the relation
$$H=\left.t^{-1}(u)\frac{d}{du}t(u)\right|_{u=0}.$$
Assuming regularity of the $R$-matrix; i.e.
$$R(0)=P$$
where $P$ is the (${\mathbb Z}_2$-graded) permutation operator,  yields
$$H=\sum_{i=1}^{L-1} h_{i(i+1)} + h_{L1}
$$
where the local two site Hamiltonians are given by
$$h=\left.\frac{d}{du}PR(u)\right|_{u=0}.$$ 

Another observable operator which is readily obtained from the 
QISM is the momentum operator which is given by 
\cite{fst} 
\be 
p=i\ln t(0) \ee 
It is more useful for our purposes to work with the 
exponentiated form 
\bea T&=& \exp (-ip) \nn \\
&=& t(0) \nn \\
&=& P_{1L}....P_{13}P_{12} \nn \eea 
which we call the translation operator. It satisfies the relations 
$$Th_{i(i+1)}T^{-1}=h_{(i+1)(i+2)},~~~~~~Th_{L1}T^{-1}=h_{12}.$$ 
Clearly 
$$[T,\,H]=0$$ 
which reflects translational invariance of the periodic model.

By using an available  construction \cite{re} for obtaining 
multiparametric quantum
algebras, it is straightforward to obtain the associated
multiparametric quantum spin chain. Below we will describe this
construction and show that in each case we can effectively 
map the additional parameters
to a ``generalized boundary condition''. 

\vspace{1cm}
~~\\
\vfil\eject 
\centerline{{\bf 3. Reshetikhin Twists}} 
~~\\

Let $(A,~\Delta,~R)$ denote a quasitriangular Hopf (super)algebra
where
$\Delta$ and $R$ denote the co-product and $R$-matrix respectively.
Suppose that there exists an element $F\in A\otimes A$ such that
\begin{eqnarray}
&&(\Delta \otimes I)(F)=F_{13}F_{23},~~~~~~\nn  \\
&&(I\otimes \Delta)(F)=F_{13}F_{12},   \nonumber    \\
&&F_{12}F_{13}F_{23}=F_{23}F_{13}F_{12}. 
\label{fyb}
\end{eqnarray}
Then $(A,~\Delta^{F},~R^F)$ is also a
quasitriangular Hopf (super)algebra with co-product 
and $R$-matrix respectively
given by
\begin{equation}
\Delta^F=F_{12}\Delta F_{12}^{-1},~~~~~~~~R^F=F_{21}RF_{12}^{-1}.
\label{df}
\end{equation}    
Throughout we refer to $F$ as a {\it twistor}.

The result stated above is a little more general than that 
originally proposed by Reshetikhin and is due to Engeldinger and Kempf 
\cite{ek}. In the original work \cite{re} Reshetikhin 
imposed the additional constraint (triangularity property)  
$$F_{12}F_{21}=I\otimes I $$ 
and in the case that $(A,~\Delta,~R)$ is an affine quantum (super)algebra
Reshetikhin gave the example  that $F$ can be chosen to be
\begin{equation}
F={\rm exp} \sum_{i<j}\left(H_i\otimes H_j-H_j\otimes
H_i\right)\phi_{ij}
\label{af}  \end{equation}
where $\{H_i\}$ is a basis for the Cartan subalgebra of the affine
quantum (super)algebra and the $\phi_{ij},~i<j$ are arbitrary complex
parameters. However, following the construction of Engeldinger and Kempf 
it is possible to choose 
\begin{equation}
F={\rm exp} \sum_{i,j}\left(H_i\otimes H_j\right)\phi_{ij} 
\label{f} \end{equation} 
which obviously gives a twistor dependent on more free parameters. 
Note that it is also possible to extend the Cartan subalgebra by
an
additional central extension (not the usual central charge) $H_0$
which
will act as a scalar multiple of the identity operator in any
irreducible representation.

Suppose that $\pi$ is a loop representation of the affine quantum
superalgebra. We let $R(u)$, $R^F(u)$ be the (super)matrix
representatives of $R$ and $R^F$ respectively, which both satisfy the
Yang-Baxter equation \reff{yb}. 
As $R(0)=P$ then $R^F(0)=P$
as a result of (\ref{df}). We may construct the transfer matrix
\begin{eqnarray}
t^F(u)&=&{\rm str}_0\left(\pi^{\otimes(L+1)}
\left(I\otimes\Delta^F_L\right)R^F_{01}\right)  \nonumber
\\
&=&{\rm str}_0\left(R^F_{0L}(u)R^F_{0(L-1)}(u)....R^F_{01}(u)
\right)
\label{tF}
\end{eqnarray}
where $\Delta^F_L$ is defined recursively through
\begin{eqnarray}
\Delta^F_L&=&\left(I\otimes I....\otimes
\Delta^F\right)\Delta^F_{L-1}
\nonumber   \\
&=&\left(\Delta^F\otimes I....\otimes I\right)\Delta^F_{L-1}
.\end{eqnarray}
Again the subscripts 0 and 1,2,...,$L$ denote the auxiliary and quantum spaces
respectively and ${\rm str}_0$ 
is the supertrace over the zeroth space.  From 
the Yang-Baxter equation it follows that the multiparametric
transfer matrices $t^F(u)$ form a commuting family.
The associated multiparametric spin chain Hamiltonian is given by
\begin{eqnarray}
H^F&=&\left.\left(t^F(u)\right)^{-1}\frac{d}{du}t^F(u)\right|_{u=0}
\nonumber  \\
&=&\sum_{i=1}^{L-1}h^F_{i,i+1}+h^F_{L1}   \label{HF} \end{eqnarray}
 with
 \bea h^F&=&\frac{d}{du}\left.PR^F(u)\right|_{u=0} \nn \\
&=&FhF^{-1}.  \label{mham} \eea 
The above construction allows us a means to incorporate arbitrary parameters
(through the $\phi_{ij}$) into the Hamiltonian without corrupting 
integrability. We will refer to such extra variables in the model as 
{\it gauge parameters}.  

Through use of (\ref{fyb}) we may alternatively write
$$t^F(u)={\rm str}_0 \left(\pi^{\otimes(L+1)}\left(I\otimes
J_L\right)
\left[(I\otimes
\Delta_L)(F_{10}R_{01}F_{10})\right]\left(I\otimes
J_L\right)^{-1}\right)$$
with
\begin{eqnarray}
J_L&=&G_{L-1}G_{L-2}....G_1,   \nonumber  \\
G_i&=&F_{iL}F_{i(L-1)}....F_{i(i+1)}.    \end{eqnarray}
We now define a new transfer matrix
\begin{eqnarray}
t(u)&=& J_L^{-1}t^F(u) J_L  \nonumber \\
&=&{\rm str}_0\left(\pi^{\otimes(L+1)}\left(I\otimes\Delta_L\right)
\left(F_{10}R_{01}F_{01}^{-1}\right)\right) \label{77}
\end{eqnarray}
where we have employed the convention to let $F$ denote both the
algebraic object and its (super)matrix representative. Through
further
use of (\ref{fyb}) we may show that
$$t(u)={\rm str}_0\left(F_{10}F_{20}....F_{L0}R_{0L}(u)R_{0(L-1)}(u)
....R_{01}(u)F_{01}^{-1}....F_{0L}^{-1}\right).$$
In this setting the operator $T$ assumes the form 
$$T=F^{-1}_{12}F^{-1}_{13}...F^{-1}_{1L}F_{21}F_{31}...F_{L1}
P_{1L}...P_{13}P_{12}  $$ 
and the associated Hamiltonian is given by
\begin{eqnarray}
H&=&\left.t^{-1}(u)\frac{d}{du} t(u)\right|_{u=0}
\nonumber  \\
&=&\sum_{i=1}^{L-1}h_{i,i+1} + h_b . \label{bham} 
\end{eqnarray} 
where
$$h_b=F^{-1}_{(L-1)L}...F_{1L}^{-1}F_{L(L-1)}...F_{L1}
h_{L1}F^{-1}_{L1}...F^{-1}_{(L-1)L}F_{1L}....F_{(L-1)L}. $$ 
It is thus apparent that the matrix $J_L$ transforms the multiparametric
Hamiltonian to one where the parameters appear only in 
a generalized boundary term
$h_b$.
Although the  
boundary term is transformed by a global operator, there are 
some important properties which should be noted that show that the 
generalized boundary interaction
 behaves as a two site operator. The first is that 
\be [h_b,\,h_{i(i+1)}]=0~~~~~\mathrm {for} ~~i\neq 1,\,L-1. 
\label{p1} \ee  
Thus if we think of the local Hamiltonians as observables then 
we can still independently  measure the boundary two site energies
and those within the bulk. Also translational invariance is 
maintained; viz. 
$$[T,\,H]=0$$ 
and more importantly 
$$Th_bT^{-1}=h_{12},~~~~~~~Th_{(L-1)L}T^{-1}=h_b$$
\be Th_{i(i+1)}T^{-1}=h_{(i+1)(i+2)}, ~~~\mathrm {for} ~~i\neq L-1.
\label{p2} \ee  
Consequently, such a model can still be interpreted as describing a closed
chain system. This situation bears close similarity with the closed 
quantum superalgebra invariant models \cite{lf97}.

~~\\
\centerline{{\bf 4. Jacobs-Cornwell Twists}}
~~\\

More recently, a new type of twisting 2-cocycle has been introduced 
by Jacobs and 
Cornwell \cite{jc} in their work on relating non-standard
quantum algebras to standard ones.
In notation as above, suppose there exists $F\in A \otimes A$ 
which satisfies the following relations 
\bea F_{12}F_{23}&=&F_{23}F_{12} \nn \\
\left(\Delta \otimes I\right)F&=&F_{23}F_{13} \nn \\
\left(I\otimes \Delta\right)F&=&F_{12}F_{13}. \nn 
\eea
Then  $(A,~\Delta^{F},~R^F)$ is also a quasi-triangular Hopf 
algebra with $\Delta^{F},~R^F$ given by \reff{df} above.  

As in the case of the Reshetikhin twists above, one may use the 
Jacobs-Cornwell twists to construct multiparametric chains which
can also be mapped to system where the extra parameters occur only
in a generalized 
 boundary interaction. For these models the transformation takes
the form 
\bea
J_L&=&G_{L-1}....G_2G_1   \nn \\
G_i&=&F_{i(i+1)}F_{i(i+2)}....F_{iL}   . \nn  
\eea 
The translation operator under this mapping becomes 
$$T=F^{-1}_{1L}F^{-1}_{1(L-1)}....F^{-1}_{12}F_{L1}....F_{21}
P_{1L}....P_{13}P_{12}$$ 
while the generalized boundary term $h_b$ in the Hamiltonian is 
$$h_b=F^{-1}_{1L}...F^{-1}_{(L-1)L}F_{L1}...F_{L(L-1)}h_{L1}
F^{-1}_{L(L-1)}...F^{-1}_{L1}F_{(L-1)L}...F_{1L}. $$
The properties (\ref{p1},\ref{p2}) also hold for the Jacobs-Cornwell
twists.

Finally, it is important to mention that the twistor 
\reff{f} qualifies
as both a 
Jacobs-Cornwell twist as well as a Reshetikhin twist. In general however,
these two classes of twistors are inequivalent (see \cite{jc}).

~~\\
\centerline{{\bf 5. Symmetric twists}}
~~\\
 
As mentioned earlier, the antisymmetrization condition \reff{af}
originally imposed by Reshetikhin can be relaxed and we can consider the
more general Cartan subalgebra twists of the form \reff{f} which
introduce more free parameters into the Hamiltonian. We will show here
however that these additional parameters simply describe local basis
transformations and as such to not have a bearing on the spectrum of the
model. 

We begin with the observation that all the twistors of the type 
\reff{f} close
to form a commutative group. 
Moreover, each twistor can be expressed as a product of
a symmetric and antisymmetric twistor through 
$$F=F^s.F^a$$ 
where 
$$F^s=(F_{12}.F_{21})^{1/2}, ~~~~~F^a=(F_{12}.F_{21}^{-1})^{1/2}. $$
The square root in the above expression is well defined given that the 
twistors of the form \reff{f} are defined in terms of the 
exponential of an operator. Given two Cartan elements $k,\,l$, let us consider a symmetric twist
given simply by 
\be F=\exp(k\ot l+l\ot k). \label{sf} \ee 
It is a straightforward exercise to show that 
$$F=\exp(\Delta(k.l)).U_1.U_2$$ 
where $U=\exp(-k.l)$ and consequently for a twistor of the type \reff{sf}
we simply have 
\be H^F=U_1U_2....U_LHU_1^{-1}U_2^{-1}....U_L^{-1} . \label{rel} \ee 
The fact that all symmetric twists arise as products of twistors of 
the form \reff{sf} allows us to conclude that the relation \reff{rel}
is true for any symmetric twistor. As a result, we do not expect that 
the Engeldinger-Kempf form of twistor \reff{f} will not introduce
more {\it dynamical} gauge parameters than twistors of the Reshetikhin form
\reff{af}.   

~~\\
\centerline{\bf 6. The quantum transfer matrix method}
~~\\
  
The algebraic Bethe ansatz method for the solution of the
multiparametric anisotropic $U$ model which will be employed below also
has an important application in the quantum transfer matrix method 
\cite{k,jks,kwz} which
we will  briefly describe here.

As a result of 
the previous discussions, we can conclude that the QISM allows us to
determine that the relation between the Hamiltonian and the transfer
matrix is of the general form
$$t(u)=T\exp[uH+o(u^2)]. $$  
Let us also define
$$ \o{t}(u)=\mathrm{str}_0\left(R_{10}(u)R_{20}(u)....R_{L0}(u)\right)
.$$
A similar calculation shows  that we may write 
$$\o{t}(u)=T^{-1}\exp[uH+o(u^2)]$$   
and thus 
$$\left(t(-\b/\L)\o{t}(-\b/\L)\right)^{\L/2}=
\exp[-\b H+o(1/\L)] $$
where $\b=1/kT$ and throughout we will assume that $\L$ is even.
Let us define two quantites $U$ and $\o{U}$ through the
relations
\bea U&=& \mathrm{tr}
\left(t(-\b/\L)\o{t}(-\b/\L)\right)^{\L/2}, \nn \\     
 \o{U}&=& \mathrm{str}
 \left(t(-\b/\L)\o{t}(-\b/\L)\right)^{\L/2}. \nn \eea 
Both $U$ and $\o{U}$ can be thought of as partition functions for a
classical two-dimensional lattice model on a torus which differ only in
boundary conditions. In the thermodynamic limit $\L\rightarrow \infty$ we
can neglect such a difference in the boundary conditions and conclude
that 
$$\lim_{\L\rightarrow \infty} \o{U}=\lim_{\L\rightarrow \infty} U =Z$$ 
where $Z$ is the partition function for the quantum system with
Hamiltonian $H$ (derived from the QISM) in the thermodynamic limit.

At this point we define the quantum transfer matrix with
inhomogeneity $\chi$ to be 
$$Q(u)=\mathrm{str}_0\left(R^{\mathrm{st}_{\L}}_{\L0}(\chi-u)
R_{0(\L-1)}(\chi+u)....R^{\mathrm{st}_2}_{20}(\chi-u)R_{01}(\chi+u)\right)  $$
which forms a commuting family $[Q(u),\,Q(v)]=0$ by virtue of the fact
that the Yang-Baxter equation (\ref{yb}) is expressible in the 
equivalent form   
$$ 
 R_{12}(u-v)R^{\mathrm{st}_{3}}_{31}(-u)R^{\mathrm{st}_{3}}_{32}(-v) 
= R^{\mathrm{st}_{3}}_{32}(-v)R^{\mathrm{st}_{3}}_{31}(-u)R_{12}(u-v)
$$ 
where throughout $\mathrm{st}_{i}$ refers to the supertransposition
taken over
the $i$th
space.
Note that we may write 
$$t(u)=\mathrm{str}_0\left(R^{\mathrm{st}_{0}}_{01}(u)
R^{\mathrm{st}_{0}}_{02}(u)....R^{\mathrm{st}_0}_{0L}(u)\right)
.$$
The quantum transfer matrix allows us to now express $Z$ as 
$$Z=\lim_{\L\rightarrow\infty} \mathrm{str}\l(Q(0)^L\r)$$ 
with $\chi=-\b/\L$. 
The thermodynamical properties are determined by the maximum eigenvalue
of $Q(u)$.
Given that $Q(u)$ forms a commuting family, the traditional Bethe
ansatz methods can be applied for the diagonalization.

~~\\ 
\centerline{\bf  7. Multiparametric anisotropic $U$ model}
~~\\

As an application of the above formalism, 
here we will introduce the multiparametric 
anisotropic $U$ model derived from  an
$R$-matrix obtained from the quantum superalgebra $U_q(sl(2|1))$.
This superalgebra has simple generators
$\{e_0,f_0,h_0
,e_1,f_1,h_1\}$ corresponding to the simple roots associated  
with the Cartan matrix
$$A=\pmatrix{0&1 \cr -1&2}.$$
It is worthwhile mentioning here that we work in the standard root
system of one bosonic and one fermionic simple root. Another system of
simple roots exists which we will not consider (see e.g. \cite{a}).
For a full definition of the algebra we refer to \cite{dglz}.

This algebra admits a non-trivial one parameter family of
four-dimensional
representations, which we label $\pi$, given by
\bea \pi(e_0)&=&\sqrt{[\a]}e^1_2+\sqrt{[\a+1]}e^3_4 \nn \\
 \pi(f_0)&=&\sqrt{[\a]}e^2_1+\sqrt{[\a+1]}e^4_3 \nn \\
 \pi(h_0)&=&\a (e^1_1+e^2_2)+(\a+1)(e^3_3+e_4^4) \nn \\
 \pi(e_1)&=&-e^2_3 \nn \\
 \pi(f_1)&=&-e^3_2 \nn  \\
 \pi(h_1)&=&e^2_2-e^3_3. \label{rep} \eea
 Above the indices of the elementary matrices
 $e^i_j$ carry the
 ${\mathbb Z}_2$-grading $(1)=(4)=0,\,(2)=(3)=1$
 and we employ the notation
 $$[x]=\frac{q^x-q^{-x}}{q-q^{-1}}.$$
 Throughout we work under the assumption that $q$ is generic.
 Associated with this representation  there is
 a solution of the
 Yang-Baxter equation which is obtained
 by solving Jimbo's equations.
 The problem of obtaining this solution  has been 
 considered in \cite{ghlz,bdgz,dglz,m,a}.
Applying the construction we described earlier yields the following
multiparametric solution of (\ref{yb}) 
\begin{footnotesize}
$$R(u)=\pmatrix{ 
R^{11}_{11}&0&0&0&|&0&0&0&0&|&0&0&0&0&|&0&0&0&0& \cr
0&R^{12}_{12}&0&0&|&R^{12}_{21}&0&0&0&|&0&0&0&0&|&0&0&0&0& \cr 
0&0&R^{13}_{13}&0&|&0&0&0&0&|&R^{13}_{31}&0&0&0&|&0&0&0&0& \cr
0&0&0&R^{14}_{14}&|&0&0&R^{14}_{23}&0&|&0&R^{14}_{32}&0&0&|&R^{14}_{41}&0&0&0& 
\cr
-&-&-&-&|&-&-&-&-&|&-&-&-&-&|&-&-&-&-& \cr 
0&R^{21}_{12}&0&0&|&R^{21}_{21}&0&0&0&|&0&0&0&0&|&0&0&0&0& \cr 
0&0&0&0&|&0&R^{22}_{22}&0&0&|&0&0&0&0&|&0&0&0&0& \cr
0&0&0&R^{23}_{14}&|&0&0&R^{23}_{23}&0&|&0&R^{23}_{32}&0&0&|&R^{23}_{41}&0&0&0& 
\cr
0&0&0&0&|&0&0&0&R^{24}_{24}&|&0&0&0&0&|&0&R^{24}_{42}&0&0& \cr
-&-&-&-&|&-&-&-&-&|&-&-&-&-&|&-&-&-&-& \cr
0&0&R^{31}_{13}&0&|&0&0&0&0&|&R^{31}_{31}&0&0&0&|&0&0&0&0& \cr 
0&0&0&R^{32}_{14}&|&0&0&R^{32}_{23}&0&|&0&R^{32}_{32}&0&0&|&R^{32}_{41}&0&0&0& 
\cr
0&0&0&0&|&0&0&0&0&|&0&0&R^{33}_{33}&0&|&0&0&0&0& \cr
0&0&0&0&|&0&0&0&0&|&0&0&0&R^{34}_{34}&|&0&0&R^{34}_{43}&0& \cr
-&-&-&-&|&-&-&-&-&|&-&-&-&-&|&-&-&-&-& \cr
0&0&0&R^{41}_{14}&|&0&0&R^{41}_{23}&0&|&0&R^{41}_{32}&0&0&|&R^{41}_{41}&0&0&0& 
\cr 
0&0&0&0&|&0&0&0&R^{42}_{24}&|&0&0&0&0&|&0&R^{42}_{42}&0&0& \cr
0&0&0&0&|&0&0&0&0&|&0&0&0&R^{43}_{34}&|&0&0&R^{43}_{43}&0& \cr
0&0&0&0&|&0&0&0&0&|&0&0&0&0&|&0&0&0&R^{44}_{44}& \cr }   
$$     
\end{footnotesize}
where the non-zero entries are given by 
\be
\begin{array}{lllllll}
R^{11}_{11}&=& \frac{[u-\a]}{[u+\a]} && 
R^{12}_{12}&=& p_1\frac{[u]}{[u+\a]} \nn \\
R^{13}_{13}&=& p_2\frac{[u]}{[u+\a]} &&
R^{14}_{14}&=& p_1p_2\frac{[u][u-1]}{[u+\a][u-\a-1]} \nn \\
^*R^{12}_{21}&=& q^u\frac{[\a]}{[u+\a]}&&
^*R^{13}_{31}&=& q^u\frac{[\a]}{[u+\a]}\nn \\
R^{14}_{41}&=& q^{2u}\frac{[\a][\a+1]}{[u+\a][u-\a-1]}&&
^*R^{14}_{23}&=& p_2p_3^{-1}q^{u-1/2}\frac{[\a]^{1/2}[\a+1]^{1/2}[u]}
	       {[u+\a][u-\a-1]}\nn\\
^*R^{14}_{32}&=& -p_1p_4^{-1}q^{u+1/2}\frac{[\a]^{1/2}[\a+1]^{1/2}[u]}
	       {[u+\a][u-\a-1]}&&
R^{21}_{21}&=& p_1^{-1}\frac{[u]}{[u+\a]} \nn \\
^*R^{22}_{22}&=& 1 &&
^*R^{23}_{23}&=& p_1^{-1}p_2p_3^{-1}p_4 \frac{[u]^2}{[u+\a][u-\a-1]}\nn \\
R^{24}_{24}&=& p_2p_3^{-1}p_4\frac{[u]}{[u-\a-1]}&&
R^{21}_{12}&=& -q^{-u}\frac{[\a]}{[u+\a]}\nn \\ 
^*R^{23}_{32}&=& \frac{2q-q^{2\a+1}-q^{-2\a-1}-q^{2u+1}+q^{2u-1}}
	       {(q^{u+\a}-q^{-u-\a})(q^{u-\a-1}-q^{-u+\a+1})}&&
R^{24}_{42}&=& q^u\frac{[\a+1]}{[u-\a-1]} \nn \\
^*R^{23}_{14}&=& -p_2p_4q^{-u+1/2}\frac{[\a]^{1/2}[\a+1]^{1/2}[u]}
	       {[u+\a][u-\a-1]}&&
^*R^{23}_{41}&=& -p_1^{-1}p_4q^{u+1/2}\frac{[\a]^{1/2}[\a+1]^{1/2}[u]}
	       {[u+\a][u-\a-1]}\nn\\
R^{31}_{31}&=& p_2^{-1}\frac{[u]}{[u+\a]} &&
^*R^{32}_{32}&=& p_1p_2^{-1}p_3p_4^{-1}\frac{[u]^2}{[u+\a][u-\a-1]}\nn \\ 
^*R^{33}_{33}&=& 1 &&
R^{34}_{34}&=& p_1p_3p_4^{-1}\frac{[u]}{[u-\a-1]}\nn \\
R^{31}_{13}&=& -q^{-u}\frac{[\a]}{[u+\a]} &&
^*R^{32}_{23}&=& \frac{2q^{-1}-q^{2\a+1}-q^{-2\a-1}+q^{-2u+1}-q^{-2u-1}}
	       {(q^{u+\a}-q^{-u-\a})(q^{u-\a-1}-q^{-u+\a+1})} \nn \\
R^{34}_{43}&=& q^u\frac{[\a+1]}{[u-\a-1]}&&
^*R^{32}_{14}&=& p_1p_3q^{-u-1/2}\frac{[\a]^{1/2}[\a+1]^{1/2}[u]}
	       {[u+\a][u-\a-1]}\nn\\
^*R^{32}_{41}&=& p_2^{-1}p_3q^{u-1/2}\frac{[\a]^{1/2}[\a+1]^{1/2}[u]}
	       {[u+\a][u-\a-1]}&&
R^{41}_{41}&=& p_1^{-1}p_2^{-1}\frac{[u][u-1]}{[u+\a][u-\a-1]}\nn \\
R^{42}_{42}&=& p_2^{-1}p_3p_4^{-1}\frac{[u]}{[u-\a-1]}&&
R^{43}_{43}&=& p_1^{-1}p_3^{-1}p_4\frac{[u]}{[u-\a-1]}\nn \\
R^{44}_{44}&=& \frac{[u+\a+1]}{[u-\a-1]}&&
R^{41}_{14}&=& q^{-2u}\frac{[\a][\a+1]}{[u+\a][u-\a-1]}\nn \\
^*R^{42}_{24}&=& -q^{-u}\frac{[\a+1]}{[u-\a-1]}&&
^*R^{43}_{34}&=& -q^{-u}\frac{[\a+1]}{[u-\a-1]}\nn \\
^*R^{41}_{23}&=& p_1^{-1}p_3^{-1}q^{-u-1/2}\frac{[\a]^{1/2}[\a+1]^{1/2}[u]}
	       {[u+\a][u-\a-1]}&&  
^*R^{41}_{32}&=& -p_2^{-1}p_4^{-1}q^{-u+1/2}\frac{[\a]^{1/2}[\a+1]^{1/2}[u]}
	       {[u+\a][u-\a-1]}\nn   \end{array}   \ee  

\vspace{1cm}
and we adopt the notation 
$$[x]=\frac{q^x-q^{-x}}{q-q^{-1}}. $$  
We remind the reader that the above $R$-matrix solves 
the Yang-Baxter equation \reff{yb} subject to the rule \reff{rule} which is
a consequence of the $U_q(sl(2|1))$ superalgebraic structure 
underlying this solution. 
The significance of the $*$ notation prefixing some matrix elements 
will be explained in a subsequent section. 

Using the above solution, we can employ the QISM to obtain an integrable 
Hamiltonian as discussed earlier. Identifying the four basis states 
in the $U_q(sl(2|1))$ representation space $V$ with the 
local electronic states through
\be v^4=\l|0\r>,~~~v^3=\l|\downarrow\r>,~~~
v^2=\l|\uparrow\r>,~~~v^1=\l|\uparrow\downarrow 
\r>\label{basis} \ee  
allows us to express the local Hamiltonians in the following form
(with convenient normalization) 
\bea 
h_{i(i+1)}&=&-p_2^{-1}p_3p_4^{-1}c_{i\upa}^{\dagger}c_{(i+1)\upa}
\l(q^{-1}p_4^2[\a+1][\a]^{-1}\r)^{\frac12 n_{i\doa}}
\l(qp_3^{-2}[\a+1][\a]^{-1}\r)^{\frac12 n_{(i+1)\doa}} +\rm{h.c.} 
\nn \\
&&
-p_1^{-1}p_3^{-1}p_4c_{i\doa}^{\dagger}c_{(i+1)\doa}
\l(qp_3^2[\a+1][\a]^{-1}\r)^{\frac12 n_{i\upa}}
\l(q^{-1}p_4^{-2}[\a+1][\a]^{-1}\r)^{\frac12 n_{(i+1)\upa}} +\rm{h.c.} 
\nn \\
&&
+[\a]^{-1}(p_1^{-1}p_2^{-1}c^{\dagger}_{i\uparrow}c^{\dagger}_{i\downarrow}
c_{(i+1)\downarrow}c_{(i+1)\uparrow}+{\rm h.c.}) 
+[\a]^{-1}( n_{i\uparrow}n_{i \downarrow}+n_{i\uparrow}n_{i \downarrow})
\nn \\ 
&&
+ q^{\a+1}( n_{i\upa}+n_{i\doa}-1)+q^{-\a-1}( n_{(i+1)\upa}+n_{(i+1)\doa}-1) 
. \label{uham} \eea 
Above we have used standard notation; the operators $c^{\dagger},\, 
c$ denote fermi creation and annihilation operators and $n$ measures 
occupation number. 
For the above operator to be hermitian, it is assumed that the
parameters $p_j,\,j=1,2,3,4$ lie on the unit circle. This results from
the fact that in
the h.c. terms one has $p_j\rightarrow p_j^{-1}$.  

An immediate feature of this Hamiltonian is that it is not spin
reflection invariant. However, it is invariant with respect to spin
reflection coupled with the interchange of parameters
\be q\leftrightarrow q^{-1},\,\, 
p_1\leftrightarrow p_2, \,\,p_3\leftrightarrow p_4. \label{inv} \ee
This invariance will manifest itself in the Bethe ansatz solutions
determined later. 

Although the above local Hamiltonian depends on the four parameters
$p_j$, we can apply the unitary transformation 
$$c^{\dagger}_{i\uparrow}\rightarrow c^{\dagger}_{i\uparrow}
(p_3p_4)^{-\frac12n_{i\doa}}, 
~~~~c^{\dagger}_{i\downarrow} \rightarrow c^{\dagger}_{i\downarrow}
(p_3p_4)^{\frac12n_{i\upa}}. $$ 
which yields the following local Hamiltonians
\bea
h_{i(i+1)}&=&-p_2^{-1}p_3p_4^{-1}c_{i\upa}^{\dagger}c_{(i+1)\upa}
\l(q^{-1}p_3^{-1}p_4[\a+1][\a]^{-1}\r)^{\frac12 n_{i\doa}}
\l(qp_3^{-1}p_4[\a+1][\a]^{-1}\r)^{\frac12 n_{(i+1)\doa}} +\rm{h.c.}
\nn \\
&&
-p_1^{-1}p_3^{-1}p_4c_{i\doa}^{\dagger}c_{(i+1)\doa}
\l(qp_3p_4^{-1}[\a+1][\a]^{-1}\r)^{\frac12 n_{i\upa}}
\l(q^{-1}p_3p_4^{-1}[\a+1][\a]^{-1}\r)^{\frac12 n_{(i+1)\upa}} +\rm{h.c.}
\nn \\
&&
+[\a]^{-1}(p_1^{-1}p_2^{-1}c^{\dagger}_{i\uparrow}c^{\dagger}_{i\downarrow}
c_{(i+1)\downarrow}c_{(i+1)\uparrow}+{\rm h.c.})
+[\a]^{-1}( n_{i\uparrow}n_{i \downarrow}+n_{i\uparrow}n_{i \downarrow})
\nn \\
&&
+ q^{\a+1}( n_{i\upa}+n_{i\doa}-1)+q^{-\a-1}(
n_{(i+1)\upa}+n_{(i+1)\doa}-1)
. \label{uham1} \eea
It is clear that there are only three independent gauge parameters
$p_1,\,p_2,\,p_3p_4^{-1}$. This result is confirmed  by the transfer matrix
eigenvalue expression and Bethe ansatz equations, which will be derived
below,
as they exhibit dependency on only these three independent gauge parameters. 
In the context of our earlier discussion, the parameter $p_3p_4$ enters
into the Hamiltonian by means of a symmetric twistor and thus may be
transformed away. 

A final comment here is that the above Hamiltonian is not the most
general that can be obtained through this procedure. By taking the 
tensor product representation of $U_q(sl(2|1))$ with different values of
the free parameter $\a$ in (\ref{rep}) a more general solution of the
Yang-Baxter equation is obtained. For the gauge free case the explicit
form is given in \cite{g99,g00}. Using the method proposed in \cite{l},
a Hamiltonian can still be derived through this solution. 

\vfil\eject
\centerline{{\bf 8. Bethe ansatz solution}}
~~\\

Having derived the model, we now use the method of the algebraic Bethe 
ansatz in order to find the eigenvalues of the transfer matrix which in 
turn permits us to calculate a formula for the energy levels of the 
Hamiltonian. Since we are working in a ``higher spin'' representation 
of $U_q(sl(2|1))$, we follow the approach first proposed for the study 
of the integrable spin 1 chain \cite{b,bt,lfk}. Rather than seek 
``creation operators'' over the pseudo-vacuum from the action of 
$R(u)$ (see e.g \cite{rm}) we instead appeal to an algebraic structure 
derived from a lower rank object. 
 In order to achieve this goal, we first introduce the 
matrices $\rr(u)\in{\rm End} (W\otimes W)$ and $L(u)\in{\rm End} (W
 \otimes V)$ where $W$ denotes a three dimensional space and $V$ is the 
four dimensional representation space of $U_q(sl(2|1))$ as before. 
We require these operators to satisfy the following forms of the 
Yang-Baxter equation 
\bea \rr_{12}(u-v)\rr_{13}(u)\rr_{23}(v)&=&\rr_{23}(v)\rr_{13} 
(u)\rr_{12}(u-v), \label{yb1} \\
\rr_{12}(u-v)L_{13}(u)L_{23}(v)&=&L_{23}(v)L_{13}(u)\rr_{12}(u-v), 
\label{yb2} \\
L_{12}(u-v)L_{13}(u)R_{23}(v)&=&R_{23}(v)L_{13}(u)L_{12}(u-v). 
\label{yb3} 
\eea

Explicitly, $\rr(u)$ and $L(u)$ satisfying the above relations 
take the form   

\begin{footnotesize}
$$\rr(u)=\pmatrix{
\rr^{11}_{11}&0&0&|&0&0&0&|&0&0&0& \cr
0&\rr^{12}_{12}&0&|&\rr^{12}_{21}&0&0&|&0&0&0& \cr
0&0&\rr^{13}_{13}&|&0&0&0&|&\rr^{13}_{31}&0&0& \cr
-&-&-&|&-&-&-&|&-&-&-& \cr
0&\rr^{21}_{12}&0&|&\rr^{21}_{21}&0&0&|&0&0&0& \cr
0&0&0&|&0&\rr^{22}_{22}&0&|&0&0&0& \cr
0&0&0&|&0&0&\rr^{23}_{23}&|&0&\rr^{23}_{32}&0& \cr
-&-&-&|&-&-&-&|&-&-&-& \cr
0&0&\rr^{31}_{13}&|&0&0&0&|&\rr^{31}_{31}&0&0& \cr
0&0&0&|&0&0&\rr^{32}_{23}&|&0&\rr^{32}_{32}&0& \cr
0&0&0&|&0&0&0&|&0&0&\rr^{33}_{33}& \cr
} $$
\end{footnotesize}
\be
\begin{array}{lllllll}
\rr^{11}_{11}&=&[u-1]   &&
\rr^{12}_{12}&=&p_1^{-1}p_2p_3^{-1}p_4[u]   \nn \\
\rr^{13}_{13}&=&p_2p_3^{-1}p_4[u]   &&
\rr^{12}_{21}&=&-q^{u}   \nn \\ 
^*\rr^{13}_{31}&=&q^{u}   &&
\rr^{21}_{21}&=&p_1p_2^{-1}p_3p_4^{-1}[u]   \nn\\
\rr^{22}_{22}&=&[u-1]   &&
\rr^{23}_{23}&=&p_1p_3p_4^{-1}[u]   \nn\\
\rr^{21}_{12}&=&-q^{-u}   &&
^*\rr^{23}_{32}&=&q^{u}   \nn\\
^*\rr^{31}_{31}&=&p_2^{-1}p_3p_4^{-1}[u]   && 
^*\rr^{32}_{32}&=&p_1^{-1}p_3^{-1}p_4[u]  \nn\\
\rr^{33}_{33}&=&[u+1]   &&
^*\rr^{31}_{13}&=&-q^{-u}   \nn\\
^*\rr^{32}_{23}&=&-q^{-u}   &&  && \nn \\
\end{array} \nn \ee 

\begin{footnotesize}
$$L(u)=\pmatrix{
L^{11}_{11}&0&0&0&|&0&0&0&0&|&0&0&0&0& \cr
0&L^{12}_{12}&0&0&|&0&0&0&0&|&0&0&0&0& \cr
0&0&L^{13}_{13}&0&|&0&L^{13}_{22}&0&0&|&L^{13}_{31}&0&0&0& \cr
0&0&0&L^{14}_{14}&|&0&0&0&0&|&0&L^{14}_{32}&0&0&
\cr
-&-&-&-&|&-&-&-&-&|&-&-&-&-& \cr
0&0&0&0&|&L^{21}_{21}&0&0&0&|&0&0&0&0& \cr
0&0&L^{22}_{13}&0&|&0&L^{22}_{22}&0&0&|&L^{22}_{31}&0&0&0& \cr
0&0&0&0&|&0&0&L^{23}_{23}&0&|&0&0&0&0& \cr
0&0&0&0&|&0&0&0&L^{24}_{24}&|&0&0&L^{24}_{33}&0& \cr
-&-&-&-&|&-&-&-&-&|&-&-&-&-& \cr
0&0&L^{31}_{13}&0&|&0&L^{31}_{22}&0&0&|&L^{31}_{31}&0&0&0& \cr
0&0&0&L^{32}_{14}&|&0&0&0&0&|&0&L^{32}_{32}&0&0& \cr
0&0&0&0&|&0&0&0&L^{33}_{24}&|&0&0&L^{33}_{33}&0& \cr
0&0&0&0&|&0&0&0&0&|&0&0&0&L^{34}_{34}& \cr
} $$
\end{footnotesize}
\be
\begin{array}{lllllll} 
L^{11}_{11}&=&p_1^{-1}[u-\a/2-1]   &&
L^{12}_{12}&=&[u-\a/2-1]   \nn \\
L^{13}_{13}&=&p_1^{-1}p_2p_3^{-1}p_4[u-\a/2]   &&
L^{14}_{14}&=&p_2p_3^{-1}p_4[u-\a/2]   \nn \\
^*L^{13}_{31}&=&-q^{u+1/2}p_1^{-1}p_4\sqrt{[\a]}  &&
L^{13}_{22}&=&-q^{u-\a/2}   \nn \\
^*L^{14}_{32}&=&q^{u}\sqrt{[\a+1]}   &&
L^{21}_{21}&=&p_2^{-1}[u-\a/2-1]   \nn \\
L^{22}_{22}&=&p_1p_2^{-1}p_3p_4^{-1}[u-\a/2]   &&
L^{23}_{23}&=&[u-\a/2-1]   \nn \\
L^{24}_{24}&=&p_1p_3p_4^{-1}[u-\a/2]   &&
L^{22}_{13}&=&-q^{-u+\a/2}   \nn \\
^*L^{22}_{31}&=&q^{u-1/2}p_2^{-1}p_3\sqrt{[\a]}   &&
^*L^{24}_{33}&=&q^{u}\sqrt{[\a+1]}   \nn \\
L^{31}_{31}&=&p_1^{-1}p_2^{-1}[u+\a/2-1]   &&
^*L^{32}_{32}&=&p_2^{-1}p_3p_4^{-1}[u+\a/2]   \nn \\
^*L^{33}_{33}&=&p_1^{-1}p_3^{-1}p_4[u+\a/2]  &&
L^{34}_{34}&=&[u+\a/2+1]   \nn \\
L^{31}_{13}&=&q^{-u-1/2}p_1^{-1}p_3^{-1}\sqrt{[\a]}  &&
L^{31}_{22}&=&-q^{-u+1/2}p_2^{-1}p_4^{-1}\sqrt{[\a]}   \nn \\
^*L^{32}_{14}&=&-q^{-u}\sqrt{[\a+1]}   &&
^*L^{33}_{24}&=&-q^{-u}\sqrt{[\a+1]}   \nn \\
\end{array} \ee

The existence of these solutions is a consequence of the representation
theory of the untwisted affine extension of $U_q(sl(2|1))$ (e.g. see 
\cite{dglz}) where the three dimensional space corresponds to the
fundamental module.
Again, multiplication of tensor products in 
(\ref{yb1},\ref{yb2},\ref{yb3}) 
is to be undertaken with consideration to (\ref{rule}).
In order the simplify the task of the Bethe ansatz approach we
work here with the solutions of (\ref{yb1},\ref{yb2},\ref{yb3}) 
without the $\mathbb{Z}_2$-graded tensor rule. This is achieved by 
a change of sign by those matrix elements of $\rr(u),\,L(u),\, 
R(u)$ prefixed with a $*$. Again, we refer to \cite{dglz} for a 
justification of this approach. We comment that in the space $W$ 
the $\mathbb{Z}_2$-grading of the basis states is $(1)=(2)=1,
\,(3)=0$. 

Let us introduce two
associative algebras $Y$ and $Z$ with elements 
$\{Y^i_j(u)\}_{i,j=1}^3$ and $\{Z^i_j
(u)\}_{i,j=1}^4$ respectively and subject to the relations
\bea \rr_{12}(u-v)Y_{13}(u)Y_{23}&=&Y_{23}(v)Y_{13}(u)\rr_{12}(u-v), 
\label{yba1} \\
R_{12}(u-v)Z_{13}(u)Z_{23}(v)&=&Z_{23}(v)Z_{13}(u)R_{12}(u-v), 
\label{yba2} \\
L_{12}(u-v)Y_{13}(u)Z_{23}(v)&=&Z_{23}(v)Y_{13}(u)L_{12}(u-v). 
\label{yba3} \eea
where 
\be Y(u)=\sum_{i,j}e^i_j\otimes Y^i_j(u),~~~Z(u)=\sum_{i,j}e^i_j\otimes
Z^i_j(u). \ee 
The associativity of these algebras is guaranteed by the relations 
\reff{yb} and \reff{yb1}. 
By comparison with eq. (\ref{yb}) we see that the monodromy matrix provides
a representation of the $Z$ algebra acting on the module $V^{\otimes L}
$ by the mapping
\be \pi\left( Z^i_j(v)\right)^k_l=(-1)^{(i)(l)+(j)(l)+(i)(k)}
T^{ik}_{jl}(v). \ee 
Moreover, the transfer matrix is expressible in terms of this
representation by 
\be t(v)=\sum_{i=1}^4(-1)^{(i)+(i)(k)}\pi\left(Z^i_i(v)\right)^k_l.
\label{grad}\ee
The phase factors present above are required since the $Z$ is defined in
terms of the non-graded $L$-operator.

We also define an auxiliary monodromy matrix through
$$U(u)=L_{0L}(u)....L_{02}(u)L_{01}(u).$$ 
Likewise, the auxiliary monodromy matrix $U(u)$ gives a representation
of the $Y$ algebra through the action 
\be \pi\left( Y^i_j(u)\right)^k_l=(-1)^{(i)(l)+(j)(l)+(i)(k)}
U^{ik}_{jl}(u) \ee
while the existence of the solution for (\ref{yb3}) permits us to 
appeal to (\ref{yba3}) for  algebraic relations between elements of
$Y$ and $Z$. 
In the following we omit the symbol $\pi$ for ease of notation.

We choose as a reference state (pseudo-vacuum) for the Bethe ansatz procedure 
the $L$-fold copy of the highest weight state in $V$; viz.
\be \Phi^0=(v^1)^{\ot L} \ee 
which itself is an eigenstate of the transfer matrix with the eigenvalue
\be \l(\frac{[v-\a]}{[v+\a]}\r)^L-\l(p_1^{-1}\frac{[v]}{[v+\a]}\r)^L 
-\l(p_2^{-1}\frac{[v]}{[v+\a]}\r)^L+\l(p_1^{-1}p_2^{-1}\frac{[v][v-1]}{
[v+\a][v-\a-1]}\r)^L. \ee 
In order to obtain more eigenstates of the transfer matrix we adopt the
ansatz 
\be \Phi^j=\sum_{\b} S^{\{\b\}}(\{u\})\Phi^0 F^j_{\{\b\}}  
\label{ansatz}\ee 
where the $F^j_{\{\b\}}$ are undetermined co-efficients and we have set
\be S^{\{\b\}}(\{u\})=Y_3^{\b_1}(u_1)Y_3^{\b_2}(u_2)....Y_3^{\b_N}(u_N) 
~~~~~~\b_i=1,2. \ee 
The motivation for this choice of ansatz is standard in the sense that a
routine calculation shows  
\be Y_i^j(u)\Phi^0=0 ~~~~\forall j\neq i \neq 3 \label{cond} \ee  
while $\Phi^0$ is an eigenstate of $Y^i_i(u),\,i=1,2,3.$ 

We now appeal to the algebraic relations given by equation (\ref{yba3}) 
in order to determine the constraints on the variables $u_i$ needed to
force equation (\ref{ansatz}) to be an eigenstate. Of the numerous relations
resulting from (\ref{yba3}) we need only consider the following.
\bea 
Z^1_1(v)Y_3^1(u)&=& p_2\frac{[u-v-\a/2-1]}{[u-v+\a/2-1]}Y_3^1(u)Z^1_1(v)
\nn \\ 
&&~~~~-q^{(u-v)}
\frac{\sqrt{[\a]}}{[u-v-\a/2-1]}\left(q^{1/2}p_2p_4Z_3^1(v)Y^1_1(u)-
q^{-1/2}p_1p_3Z_2^1(v)Y_2^1(u)\right) \nn \\
Z^1_1(v)Y_3^2(u)&=& p_1\frac{[u-v-\a/2-1]}{[u-v+\a/2-1]}Y_3^2(u)Z^1_1(v)
\nn \\ &&~~~~-q^{(u-v)}
\frac{\sqrt{[\a]}}{[u-v-\a/2-1]}\left(q^{1/2}p_2p_4Z_3^1(v)Y_1^2(u)- 
q^{-1/2}p_1p_3Z_2^1(v)Y^2_2(u)\right) \nn \\  
Z^4_4(v)Y^1_3(u)&=&p_2p_3^{-1}p_4\frac{[u-v-\a/2-1]}{[u-v+\a/2]}Y^1_3(u)
Z^4_4(v)-q^{(u-v)}p_2p_3^{-1}p_4\frac{\sqrt{[\a+1]}}{[u-v+\a/2]}
Z^2_4(v)Y^3_3(u) \nn \\
Z^4_4(v)Y^2_3(u)&=&p_1p_3p_4^{-1}\frac{[u-v-\a/2-1]}{[u-v+\a/2]}Y^2_3(u)
Z^4_4(v)-q^{(u-v)}p_1p_3p_4^{-1}\frac{\sqrt{[\a+1]}}{[u-v+\a/2]}
Z^3_4(v)Y^3_3(u) \nn \\
Z^2_2(v)Y^1_3(u)&=&-p_2p_3^{-1}p_4\frac{[u-v-\a/2-1]}{[u-v+\a/2]}
Y^1_3(u)Z^2_2(v)-q^{(u-v)}p_2p_3^{-1}p_4\frac{\sqrt{[\a+1]}}{[u-v+\a/2]}
Z^2_4(v)Y^1_1(u)\nn\\
Z^3_3(v)Y^2_3(u)&=&-p_1p_3p_4^{-1}\frac{[u-v-\a/2-1]}{[u-v+\a/2]}
Y^2_3(u)Z^3_3(v)-q^{(u-v)}p_1p_3p_4^{-1}\frac{\sqrt{[\a+1]}}{[u-v+\a/2]}
Z^3_4(v)Y^2_2(u)\nn\\
Z^2_3(v)Y^1_3(u)&=&-p_1p_3p_4^{-1}\frac{[u-v-\a/2-1]}{[u-v+\a/2]}
Y^1_3(u)Z^2_3(v)-q^{(u-v)}p_1p_3p_4^{-1}\frac{\sqrt{[\a+1]}}{[u-v+\a/2]}
Z^2_4(v)Y^1_2(u)\nn\\
Z^3_2(v)Y^2_3(u)&=&-p_2p_3^{-1}p_4\frac{[u-v-\a/2-1]}{[u-v+\a/2]}
Y^2_3(u)Z^3_2(v)-q^{(u-v)}p_2p_3^{-1}p_4\frac{\sqrt{[\a+1]}}{[u-v+\a/2]}
Z^3_4(v)Y^2_1(u)\nn\\
Z^2_2(v)Y^2_3(u)&=&-p_1\frac{[u-v-\a/2-1]}{[u-v+\a/2-1]}Y^2_3(u)Z^2_2(v)
\nn \\
&&~+p_2p_3^{-1}p_4q^{(v-u-\a/2)}\frac{[u-v-\a/2-1]}{[u-v+\a/2][u-v+\a/2-1]}
Y^1_3(u)Z^3_2(v)\nn \\ 
&&~+p_2p_3^{-1}p_4q^{(u-v)}\frac{\sqrt{[\a+1]}}{
[u-v+\a/2]}Z^2_4(v)Y^2_1(u) \nn \\  
&&~-p_1p_3q^{(u-v-1/2)}\frac{\sqrt{[\a]}} 
{[u-v+\a/2-1]}Z^1_2(v)Y^3_3(u)\nn \\
&&~-p_1p_2p_4q^{(2u-2v-1/2)}\frac{\sqrt{[\a]}
\sqrt{[\a+1]}}{[u-v+\a/2-1][u-v+\a/2]}Z^1_4(v)Y_1^3(u) \nn \\ 
Z^3_3(v)Y^1_3(u)&=&-p_2\frac{[u-v-\a/2-1]}{[u-v+\a/2-1]}Y^1_3(u)Z^3_3(v)
\nn \\
&&~+p_1p_3p_4^{-1}q^{(u-v+\a/2)}\frac{[u-v-\a/2-1]}{[u-v+\a/2][u-v+\a/2-1]} 
Y^2_3(u)Z^2_3(v)\nn \\
&&~+p_1p_3p_4^{-1}q^{(u-v)}\frac{\sqrt{[\a+1]}}{
[u-v+\a/2]}Z^3_4(v)Y^1_2(u)\nn \\
&&~+p_2p_4q^{(u-v+1/2)}\frac{\sqrt{[\a]}}
{[u-v+\a/2-1]}Z^1_3(v)Y^3_3(u)\nn \\
&&~+p_1p_2p_3q^{(2u-2v+1/2)}\frac{\sqrt{[\a]}
\sqrt{[\a+1]}}{[u-v+\a/2-1][u-v+\a/2]}Z^1_4(v)Y_2^3(u) \nn \\
Z^2_3(v)Y^2_3(u)&=&-p_1^2p_2^{-1}p_3^2p_4^{-2}\frac{[u-v-\a/2-1]}{[u-v+\a/2-1]}
Y^2_3(u)Z^2_3(v)
\nn \\
&&~+p_1p_3p_4^{-1}q^{(v-u-\a/2)}\frac{[u-v-\a/2-1]}{[u-v+\a/2][u-v+\a/2-1]}
Y^1_3(u)Z^3_3(v)\nn \\
&&~-p_1p_3p_4^{-1}q^{(u-v)}\frac{\sqrt{[\a+1]}}{
[u-v+\a/2]}Z^2_4(v)Y^2_2(u)\nn \\
&&~-p_1p_3q^{(u-v-1/2)}\frac{\sqrt{[\a]}}
{[u-v+\a/2-1]}Z^1_3(v)Y^3_3(u)\nn \\
&&~-p_1^{-2}p_3^2p_4^{-1}q^{(2u-2v-1/2)}\frac{\sqrt{[\a]}
\sqrt{[\a+1]}}{[u-v+\a/2-1][u-v+\a/2]}Z^1_4(v)Y_2^3(u) \nn \\
Z^3_2(v)Y^1_3(u)&=&-p_1^{-1}p_2^2p_3^{-2}p_4^{2}
\frac{[u-v-\a/2-1]}{[u-v+\a/2-1]}
Y^1_3(u)Z^3_2(v)
\nn \\
&&~+p_2p_3^{-1}p_4q^{(u-v+\a/2)}\frac{[u-v-\a/2-1]}{[u-v+\a/2][u-v+\a/2-1]}
Y^2_3(u)Z^2_2(v)\nn \\
&&~-p_2p_3^{-1}p_4q^{(u-v)}\frac{\sqrt{[\a+1]}}{
[u-v+\a/2]}Z^3_4(v)Y^1_1(u)\nn \\
&&~+p_2p_4q^{(u-v+1/2)}\frac{\sqrt{[\a]}}
{[u-v+\a/2-1]}Z^1_2(v)Y^3_3(u)\nn \\
&&~+p_2^{-2}p_3^{-1}p_4^{2}q^{(2u-2v+1/2)}\frac{\sqrt{[\a]}
\sqrt{[\a+1]}}{[u-v+\a/2-1][u-v+\a/2]}Z^1_4(v)Y_1^3(u) \nn \\
&&
\nn\eea

In the above commutation relations two types of terms arise when the
actions of the operators $Z^1_1(v),\,Z^4_4(v)$ are evaluated on 
(\ref{ansatz}). These
terms are those giving a vector proportional to $\Phi^j$ which we call
{\it wanted terms} while all other types are {\it unwanted terms}
(u.t.). The operators in the above expressions contributing to the 
unwanted terms are all those of the 
form $Z^i_j(v)Y^k_l(u)$ with the necessity that $i< j$.  
Note that for any $Y^k_l(u)$ term which arises where
$l\neq 3$ and $k\neq 1,\,2$  we can appeal to the commutation relations
(\ref{yba1}) to commute it through to the reference state $\Phi^0$ where it
will act as a scalar. Repeated use of this procedure shows that all unwanted
terms are ultimately expressible in the form 
$$S^{\{\gamma\}}(\{u'\})Z^i_j(v)S^{\{\gamma'\}}(\{u''\})\Phi^0$$
with $i< j$ and $v$ {\it arbitrary} and so can never give a vector
proportional to $\Phi^j$. 

In terms of the fermionic representation (\ref{basis}), it is deduced
that the reference state $\Phi^0$ is the completely filled state with
two electrons of opposite spin at each lattice site.
Furthermore, it is not difficult to see that the operator $L^1_3(v)$
annihilates a spin up electron from the system while $L^2_3(v)$
annihilates a spin down electron. Using the condition that the states
(\ref{ansatz}) have well defined electron number and spin (this can be
interpreted as the conservation condition in the sense of Schultz
\cite{schultz}), we have 
\bea Z^1_1(v)\Phi^j&=&p_1^{L-n_-}p_2^{L-n_+}
\frac{[v-\a]^L}{[v+\a]^L}\prod_{i=1}^{2L-N} 
\frac{[u_i-v-\a/2-1]}{[u_i-v+\a/2-1]}\Phi^j+ {\rm u.t.} \nn \\
 Z^4_4(v)\Phi^j&=&p_1^{-n_-}p_2^{-n_+}p_3^{(n_+-n_-)}p_4^{(n_--n_+)}
\frac{[v]^L[v-1]^L}{[v+\a]^L[v-\a-1]^L} \prod_{i=1}^{2L-N} 
\frac{[u_i-v-\a/2-1]}{[u_i-v+\a/2]}\Phi^j+ {\rm u.t.} \nn \eea  
where $n_+$ and $n_-$ are respectively the number of spin up and down
electrons in the state and $N=n_++n_-$.   

The situation for determining the action of $Z^2_2(v),\,Z^3_3(v)$ is a
little more complicated. For these cases more than one wanted term may
arise in the calculation, which forces us to employ a nested Bethe
ansatz approach. 
In terms of the six-vertex  solution of the Yang-Baxter
equation with gauge parameter $l$     
\be r(v,l)=\pmatrix{[v+1]&0&0&0 \cr 0&l[v]& q^{v}&0 \cr
0&q^{-v}& l^{-1}[v] &0 \cr 0&0&0& [v+1] }
\ee 
we define the monodromy matrix 
$$t(v,\{u\},l)=r_{01}(v-u_1,l)r_{02}(v-u_2,l)....r_{0N}(v-u_N,l).$$ 
It turns out that in this
notation we may write 
\bea Z^2_2(v)\Phi^j&=&\l(p_1^{-1}\frac{[v]}{[v+\a]}\r)^L
\prod_{i=1}^{2L-N}\l(p_2p_3^{-1}p_4\frac{[u_i-v-\a/2-1]}
{[u_i-v+\a/2][u_i-v+\a/2-1]}\r) 
\nn \\
&&~~\times 
S^{\{\b'\}}(\{u\})t^{1\{\b\}}_{1\{\b'\}}
(v-\a/2,\{u\}, p_1p_2^{-1}p_3p_4^{-1})\Phi^0F^j_{
\{\b\}} + {\rm u.t.} \nn \\
Z^3_3(v)\Phi^j&=&\l(p_2^{-1}\frac{[v]}{[v+\a]}\r)^L
\prod_{i=1}^{2L-N}\l(p_1p_3p_4^{-1}\frac{[u_i-v-\a/2-1]}{[u_i-v+\a/2]
[u_i-v+\a/2-1]}\r) 
\nn \\ &&~~\times S^{\{\b'\}}(\{u\})t^{2\{\b\}}_{2\{\b'\}}(v-\a/2,
\{u\}, p_1p_2^{-1}p_3p_4^{-1})\Phi^0F^j_{
\{\b\}} + {\rm u.t.} \nn \eea 

Completing the diagonalization of the transfer matrix is
accomplished by diagonalizing the operator
\be \mu_1 t^1_1(v,\{u\},l)+\mu_2 t^2_2(v,\{u\},l) \label{6tm} \ee  
which is the transfer matrix for the six-vertex model with gauge
parameter $l$, inhomogeneities $u_i$ and twisted boundary conditions. 
This is a standard calculation 
so we simply present the eigenvalue
expression which reads 
$$\lambda(v)=\mu_1l^M\prod_{i=1}^{2L-N}[v-u_i+1]\prod_{j=1}^M\frac{ [v-w_j-1]
}{[v-w_j]}+\mu_2l^{(M+N-2L)}
\prod_{i=1}^{2L-N}[v-u_i]\prod_{j=1}^M \frac{[v-w_j+1]}{
[v-w_j]}.$$ 
Making the substitutions
\bea &&v\rightarrow v-\a/2, \nn \\
&&l\rightarrow p_1p_2^{-1}p_3p_4^{-1} \nn \\
&&\mu_1\rightarrow p_1^{-L}p_2^Np_3^{-N}p_4^N \nn \\
&&\mu_2\rightarrow p_1^Np_2^{-L}p_3^Np_4^{-N} \nn \eea 
allows us to determine the contribution to the eigenvalue expression for
the transfer matrix 
coming from the terms $Z^2_2(v)$ and $Z^3_3(v)$.  At this point
we would like to remark that the reference state used in the Bethe
ansatz diagonalization of the six-vertex transfer matrix 
(\ref{6tm}) corresponds to the state
$$L^1_3(u_1)L^1_3(u_2)....L^1_3(u_N)\Phi^0$$ 
which is an $2L-N$ electron state containing $L$  spin down electrons
and $L-N$ spin up electrons.
We can immediately deduce that $M=L-n_-$.   
Collecting these results and making the substitutions
$u_i\rightarrow u_i+1,\, w_j\rightarrow w_j+1/2$, 
we may now give the eigenvalue expression 
\bea \Lambda_1(v,u_i,w_j)&=&
p_1^{-n_-}p_2^{-n_+}p_3^{(n_+-n_-)}p_4^{(n_--n_+)}
\frac{[v]^L[v-1]^L}{[v+\a]^L[v-\a-1]^L} \prod_{i=1}^{2L-N}
\frac{[u_i-v-\a/2]}{[u_i-v+\a/2+1]} \nn \\
&&-\l(p_1^{-1}\frac{[v]}{[v+\a]}\r)^L
\prod_{i=1}^{2L-N}\l(\frac{p_2p_4[u_i-v-\a/2]}
{p_3[u_i-v+\a/2+1]}\r)\prod_{j=1}^{L-n_-}\l(\frac{p_1p_3}{p_2p_4}
\frac{[w_j-v+\a/2+3/2]}{[w_j-v+\a/2+1/2]}\r) \nn \\
&&-\l(p_2^{-1}\frac{[v]}{[v+\a]}\r)^L
\prod_{i=1}^{2L-N}\l(p_2\frac{[u_i-v-\a/2]}{
[u_i-v+\a/2]}\r) \prod_{j=1}^{L-n_-}\l(\frac{p_1p_3}{p_2p_4}
\frac{[w_j-v+\a/2-1/2]}{[w_j-v+\a/2+1/2]}\r) 
\nn \\ 
&&+p_1^{L-n_-}p_2^{L-n_+}
\frac{[v-\a]^L}{[v+\a]^L}\prod_{i=1}^{2L-N}
\frac{[u_i-v-\a/2]}{[u_i-v+\a/2]} \label{eig1}  \eea 
for the transfer matrix (\ref{tm}). 
Constraints on the parameters $u_i,\,w_j$ are given by the Bethe Ansatz
equations. In principle, these may be derived from the requirement that
all the unwanted terms cancel in the above calculation. 
A more practical way to obtain them is
to use the fact that the eigenvalue expression (\ref{eig1}) is an analytic
function of the variable $v$. Imposing this condition leads us to the
set of equations 
\bea p_2^{L}p_3^{n_--L}p_4^{L-n_-}\l(\frac{[u_i-\a/2]}{[u_i+\a/2]}\r)^L&=& 
\prod_{j=1}^{L-n_-}\frac{[u_i-w_j+1/2]}{[u_i-w_j-1/2]},~~~i=1,2,..., 2L-N \nn
\\
p_1^{-L}p_2^Lp_3^{N-2L}p_4^{2L-N}\prod_{i=1}^{2L-N}
\frac{[u_i-w_j-1/2]}{[u_i-w_j+1/2]}
&=&-\prod_{k=1}^{L-n_-}\frac{[w_j-w_k+1]}{[w_j-w_k-1]},~~~j=1,2,...,L-n_- 
\label{bae} \eea 

Alternatively, we can use the following reference state
$$L^2_3(u_1)L^2_3(u_2)....L^2_3(u_N)\Phi^0$$  
for the diagonalization of the operator (\ref{6tm}). In this instance we
obtain 
\bea \oL_1(v,u_i,w_j)&=&
p_1^{-n_-}p_2^{-n_+}p_3^{(n_+-n_-)}p_4^{(n_--n_+)}
\frac{[v]^L[v-1]^L}{[v+\a]^L[v-\a-1]^L} \prod_{i=1}^{2L-N}
\frac{[u_i-v-\a/2]}{[u_i-v+\a/2+1]} \nn \\
&&-\l(p_2^{-1}\frac{[v]}{[v+\a]}\r)^L
\prod_{i=1}^{2L-N}\l(\frac{p_1p_3[u_i-v-\a/2]}
{p_4[u_i-v+\a/2+1]}\r)\prod_{j=1}^{L-n_+}\l(\frac{p_2p_4}{p_1p_3}
\frac{[w_j-v+\a/2+3/2]}{[w_j-v+\a/2+1/2]}\r) \nn \\
&&-\l(p_1^{-1}\frac{[v]}{[v+\a]}\r)^L
\prod_{i=1}^{2L-N}\l(p_1\frac{[u_i-v-\a/2]}{
[u_i-v+\a/2]}\r) \prod_{j=1}^{L-n_+}\l(\frac{p_2p_4}{p_1p_3}
\frac{[w_j-v+\a/2-1/2]}{[w_j-v+\a/2+1/2]}\r)
\nn \\
&&+p_1^{L-n_-}p_2^{L-n_+}
\frac{[v-\a]^L}{[v+\a]^L}\prod_{i=1}^{2L-N}
\frac{[u_i-v-\a/2]}{[u_i-v+\a/2]} \label{oeig1}  \eea
for the eigenvalues of the transfer matrix. It is clear that
(\ref{oeig1}) is obtained from (\ref{eig1}) by the change of parameters
(\ref{inv}) and spin reflection, a consequence of the invariance of the
Hamiltonian (\ref{uham}) with respect to this operation. For this case the
Bethe ansatz equations assume the form
\bea
p_1^{L}p_3^{L-n_+}p_4^{n_+-L}\l(\frac{[u_i-\a/2]}{[u_i+\a/2]}\r)^L&=&
\prod_{j=1}^{L-n_+}\frac{[u_i-w_j+1/2]}{[u_i-w_j-1/2]},~~~i=1,2,...,
2L-N \nn
\\
p_1^{L}p_2^{-L}p_3^{2L-N}p_4^{N-2L}\prod_{i=1}^{2L-N}
\frac{[u_i-w_j-1/2]}{[u_i-w_j+1/2]}
&=&-\prod_{k=1}^{L-n_+}\frac{[w_j-w_k+1]}{[w_j-w_k-1]},~~~j=1,2,...,L-n_+
\nn \eea
It is not clear whether the two eigenvalue expressions presented above,
either individually or together, give the entire spectrum for the model.
No highest weight theorem for the Bethe states is applicable here due  
to the fact that the imposition of periodic boundary conditions 
does not preserve invariance with respect to the action of the elements
of $U_q(sl(2|1))$. In the $q\rightarrow 1$ limit  the invariance is restored
and one can show that the Bethe states are indeed highest weight states
similar to the methods used for other $sl(2|1)$ invariant models 
\cite{fk,flt}. However the combinatorial arguments given in those
works do not extend in any obvious manner for this case 
and to our knowledge the 
completeness of the Bethe states for the $U$ model remains an open
question.

Using the transfer matrix eigenvalue expressions 
(\ref{eig1},\ref{oeig1}) the energies of
the Bethe states in both cases are obtained through

\bea E&=&-\frac{q^{\a+1}-q^{-\a-1}}{2\ln
q}\left.\La^{-1}\frac{d\La}{du}\right|_{u=0} \nn \\ 
&=&\frac{(q^{\a}+q^{-\a})[\a+1]L}{[\a]}+\sum_{i=1}^{2L-N} 
\frac{[\a][\a+1]}{[u_i+\a/2][u_i-\a/2]}. \nn \eea 

One can alternatively use a tensor product of lowest weight states
as the pseudo-vacuum for the Bethe ansatz calculation; viz.
$$\Phi^0=(v^4)^{\ot L}.$$ 
This leads to a different form for the transfer matrix eigenvalues 
and Bethe ansatz equations. The procedure used is analogous to that 
above, and so we will omit the details of the calculation and just 
give the final results. The transfer matrix eigenvalues are  
\bea \Lambda_2(v,u_i,w_j)&=&
p_1^{L-n_+}p_2^{L-n_-}
\frac{[v]^L[v-1]^L}{[v+\a]^L[v-\a-1]^L} \prod_{i=1}^{N}
\frac{[u_i-v+\a/2+1/2]}{[u_i-v-\a/2+1/2]} \nn \\
&&-\l(\frac{p_2p_4}{p_3}\frac{[v]}{[v-\a-1]}\r)^L
\prod_{i=1}^{N}\l(p_1^{-1}\frac{[u_i-v+\a/2+1/2]}
{[u_i-v-\a/2+1/2]}\r)\prod_{j=1}^{n_+}\l(\frac{p_1p_3}{p_2p_4}
\frac{[w_j-v-\a/2+1]}{[w_j-v-\a/2]}\r) \nn \\
&&-\l(\frac{p_1p_3[v]}{p_4[v-\a-1]}\r)^L
\prod_{i=1}^{N}\l(\frac{p_4[u_i-v+\a/2+1/2]}{
p_1p_3[u_i-v-\a/2-1/2]}\r) \prod_{j=1}^{n_+}\l(\frac{p_1p_3}{p_2p_4}
\frac{[w_j-v-\a/2-1]}{[w_j-v-\a/2]}\r)
\nn \\
&&+p_1^{-n_+}p_2^{-n_-}p_3^{(n_--n_+)}p_4^{(n_+-n_-)}
\frac{[v+\a+1]^L}{[v-\a-1]^L}\prod_{i=1}^{N}
\frac{[u_i-v+\a/2+1/2]}{[u_i-v-\a/2-1/2]} \label{eig2}  \eea
where the Bethe ansatz equations read 
\bea
p_1^{L}p_3^{L-n_+}p_4^{n_+-L}\l(\frac{[u_i-\a/2-1/2]}{[u_i+\a/2+1/2]}\r)^L&=&
\prod_{j=1}^{n_+}\frac{[u_i-w_j-1/2]}{[u_i-w_j+1/2]},~~~i=1,2,..., N
\nn
\\
p_1^{-L}p_2^{L}p_3^{N-2L}p_4^{2L-N}\prod_{i=1}^{N}
\frac{[u_i-w_j-1/2]}{[u_i-w_j+1/2]}
&=&-\prod_{k=1}^{n_+}\frac{[w_j-w_k+1]}{[w_j-w_k-1]},~~~j=1,2,...,n_- 
.\nn \eea
Just as in the previous case there is a second expression for the 
eigenvalues, due to a different choice of reference state in the nesting,
which is 
\bea \oL_2(v,u_i,w_j)&=&
p_1^{L-n_+}p_2^{L-n_-}
\frac{[v]^L[v-1]^L}{[v+\a]^L[v-\a-1]^L} \prod_{i=1}^{N}
\frac{[u_i-v+\a/2+1/2]}{[u_i-v-\a/2+1/2]} \nn \\
&&-\l(\frac{p_1p_3}{p_4}\frac{[v]}{[v-\a-1]}\r)^L
\prod_{i=1}^{N}\l(p_2^{-1}\frac{[u_i-v+\a/2+1/2]}
{[u_i-v-\a/2+1/2]}\r)\prod_{j=1}^{n_-}\l(\frac{p_2p_4}{p_1p_3}
\frac{[w_j-v-\a/2+1]}{[w_j-v-\a/2]}\r) \nn \\
&&-\l(\frac{p_2p_4[v]}{p_3[v-\a-1]}\r)^L
\prod_{i=1}^{N}\l(\frac{p_3[u_i-v+\a/2+1/2]}{
p_2p_4[u_i-v-\a/2-1/2]}\r) \prod_{j=1}^{n_-}\l(\frac{p_2p_4}{p_1p_3}
\frac{[w_j-v-\a/2-1]}{[w_j-v-\a/2]}\r)
\nn \\
&&+p_1^{-n_+}p_2^{-n_-}p_3^{(n_--n_+)}p_4^{(n_+-n_-)}    
\frac{[v+\a+1]^L}{[v-\a-1]^L}\prod_{i=1}^{N}
\frac{[u_i-v+\a/2+1/2]}{[u_i-v-\a/2-1/2]}. \label{oeig2}  \eea
For this expression, the Bethe ansatz equations are
\bea
p_2^{L}p_3^{n_--L}p_4^{L-n_-}\l(\frac{[u_i-\a/2-1/2]}{[u_i+\a/2+1/2]}\r)^L&=&
\prod_{j=1}^{n_-}\frac{[u_i-w_j-1/2]}{[u_i-w_j+1/2]},~~~i=1,2,..., N
\nn
\\
p_1^{L}p_2^{-L}p_3^{2L-N}p_4^{N-2L}\prod_{i=1}^{N}
\frac{[u_i-w_j-1/2]}{[u_i-w_j+1/2]}
&=&-\prod_{k=1}^{n_-}\frac{[w_j-w_k+1]}{[w_j-w_k-1]},~~~j=1,2,...,n_+
.\nn \eea
One can see that the formulae $\Lambda_2(v,u_i,w_j)$ and 
$\oL_2(v,u_i,w_j)$ are related through the mapping (\ref{inv}) and spin 
reflection as in the previous example. Moreover, there is a mapping from
 $\Lambda_1(v,u_i,w_j)$ to 
 $\Lambda_2(v,u_i,w_j)$ and  
$\oL_1(v,u_i,w_j)$ to 
$\oL_2(v,u_i,w_j)$ (and the corresponding Bethe ansatz equations) 
which is given by 
\be p_1\rightarrow p_2^{-1}p_3p_4^{-1},\,p_2\rightarrow
p_1^{-1}p_3^{-1}p_4,\,p_3\rightarrow p_4,\, p_4\rightarrow p_3,\,
\a\rightarrow -
\a-1,\, n_+\rightarrow L-n_-, \, n_-\rightarrow L-n_+. \label{inv1} \ee 
The existence of this relationship can be understood in terms of the dual 
representation of (\ref{rep}). Recall that for any 
representation of a quantum superalgebra $\pi$ the dual representation 
$\pi^*$ is defined for each element $a$ in the superalgebra through
$$\pi^*(a)=\pi(S(a))^{\rm st}$$ 
where st denotes the supertransposition as before 
and $S$ is the antipode. For 
the representation (\ref{rep}) under consideration here, the dual 
representation is equivalent (i.e. up to a basis transformation) to the 
change of variable $\a\rightarrow -\a-1$. The change in the gauge 
parameters is also a result of the dualization of the Cartan elements
in the definition of (\ref{f}).  
 
The transfer matrix eigenvalue expressions 
(\ref{eig2},\ref{oeig2}) for this second form of the Bethe ansatz 
give the energies of
the Bethe states  through
\bea E&=&-\frac{q^{\a+1}-q^{-\a-1}}{2\ln
q}\left.\La^{-1}\frac{d\La}{du}\right|_{u=0} \nn \\
&=&-(q^{\a+1}+q^{-\a-1})L-\sum_{i=1}^{N}
\frac{[\a+1]^2}{[u_i+\a/2+1/2][u_i-\a/2-1/2]}. \nn \eea

A third form of the Bethe ansatz also exists for this model,
 which was discovered
by Pfanm\"uller and Frahm \cite{pf97} in the isotropic case ($q=1$)
with all gauge parameters equal to 1. The extension to the present model
is achieved by a similar construction and so we will again omit the details
and refer the interested reader to \cite{pf97} for the methodology. Again,
there are two types of eigenvalue expression which are  
\bea &&\Lambda_3(v,u_i,w_j) \nn \\
&&=
\l(\frac{p_2[v]}{[v+\a]}\r)^L \prod_{i=1}^{L-n_-}
\frac{p_1[u_i-v-\a/2+1/2]}{[u_i-v+\a/2+1/2]} 
\prod_{j=1}^{n_+}\frac{[w_j-v+\a/2]}{p_2[w_j-v-\a/2]}\nn \\
~~&&+\l(\frac{p_4[v]}{p_1p_3[v-\a-1]}\r)^L
\prod_{i=1}^{L-n_-}\l(\frac{p_1p_3[u_i-v-\a/2-1/2]}
{p_4[u_i-v+\a/2+1/2]}\r)\prod_{j=1}^{n_+}\l(
\frac{p_3[w_j-v+\a/2+1]}{p_2p_4[w_j-v-\a/2]}\r) \nn \\
~~&&-\l(\frac{p_2p_4[v]^2}{p_1p_3[v+\a][v-\a-1]}\r)^L
\prod_{i=1}^{L-n_-}\l(\frac{p_1[u_i-v-\a/2+1/2]}{
[u_i-v+\a/2+1/2]}\r) \prod_{j=1}^{n_+}\l(
\frac{p_3[w_j-v+\a/2+1]}{p_2p_4[w_j-v-\a/2]}\r)
\nn \\
~~&&- \prod_{i=1}^{L-n_-} \frac{p_1p_3[u_i-v-\a/2-1/2]}{p_4[u_i-v+\a/2+1/2]}
\prod_{j=1}^{n_+} \frac{[w_j-v+\a/2]}{p_2[w_j-v-\a/2]} \label{eig3}  \eea
subject to the Bethe ansatz equations
\bea
p_1^{-L}p_3^{n_+-L}p_4^{L-n_+}\l(\frac{[u_i+\a/2+1/2]}{[u_i-\a/2-1/2]}\r)^L&=&
\prod_{j=1}^{n_+}\frac{[w_j-u_i-1/2]}{[w_j-u_i+1/2]} \nn \\
 p_2^{L}p_3^{n_--L}p_4^{L-n_-}\l(\frac{[w_j-\a/2]}{[w_j+\a/2]}\r)^L&=&
\prod_{i=1}^{L-n_-}\frac{[u_i-w_j-1/2]}{[u_i-w_j+1/2]} 
\label{bae3} \eea 
for which the energy is given by 
$$E=\sum_{i=1}^{L-n_-} \frac{[\a+1]^2}{[u_i+\a/2+1/2][u_i-\a/2-1/2]}
-\sum_{j=1}^{n_+}\frac{[\a][\a+1]}{[w_j+\a/2][w_j-\a/2]}$$
and 
\bea &&\oL_3(v,u_i,w_j) \nn \\
&=&
\l(\frac{p_1[v]}{[v+\a]}\r)^L \prod_{i=1}^{L-n_+}
\frac{p_2[u_i-v-\a/2+1/2]}{[u_i-v+\a/2+1/2]}
\prod_{j=1}^{n_-}\frac{[w_j-v+\a/2]}{p_1[w_j-v-\a/2]}\nn \\
&&+\l(\frac{p_3[v]}{p_2p_4[v-\a-1]}\r)^L
\prod_{i=1}^{L-n_+}\l(\frac{p_2p_4[u_i-v-\a/2-1/2]}
{p_3[u_i-v+\a/2+1/2]}\r)\prod_{j=1}^{n_-}\l(
\frac{p_4[w_j-v+\a/2+1]}{p_1p_3[w_j-v-\a/2]}\r) \nn \\
&&-\l(\frac{p_1p_3[v]^2}{p_2p_4[v+\a][v-\a-1]}\r)^L
\prod_{i=1}^{L-n_+}\l(\frac{p_2[u_i-v-\a/2+1/2]}{
[u_i-v+\a/2+1/2]}\r) \prod_{j=1}^{n_-}\l(
\frac{p_4[w_j-v+\a/2+1]}{p_1p_3[w_j-v-\a/2]}\r)
\nn \\
&&- \prod_{i=1}^{L-n_+}
\frac{p_2p_4[u_i-v-\a/2-1/2]}{p_3[u_i-v+\a/2+1/2]}
\prod_{j=1}^{n_-} \frac{[w_j-v+\a/2]}{p_1[w_j-v-\a/2]} \label{oeig3}
\eea
in which case the Bethe ansatz equations are
\bea
p_2^{-L}p_3^{L-n_-}p_4^{n_--L}\l(\frac{[u_i+\a/2+1/2]}{[u_i-\a/2-1/2]}\r)^L&=&
\prod_{j=1}^{n_-}\frac{[w_j-u_i-1/2]}{[w_j-u_i+1/2]} \nn \\
 p_2^{L}p_3^{n_+-L}p_4^{L-n_+}\l(\frac{[w_j-\a/2]}{[w_j+\a/2]}\r)^L&=&
 \prod_{i=1}^{L-n_+}\frac{[u_i-w_j-1/2]}{[u_i-w_j+1/2]} 
\label{obae3} \eea
and the energy expression is 
$$E=\sum_{i=1}^{L-n_+} \frac{[\a+1]^2}{[u_i+\a/2+1/2][u_i-\a/2-1/2]}
-\sum_{j=1}^{n_-}\frac{[\a][\a+1]}{[w_j+\a/2][w_j-\a/2]}.$$ 
Again, the two eigenvalue expressions and associated 
Bethe ansatz equations are related through spin 
reflection coupled with the change of parameters 
(\ref{inv}). A more striking feature is that 
(\ref{eig3},\ref{bae3},\ref{oeig3},\ref{obae3})
 are all invariant under (\ref{inv1}) with an accompanying interchange
of the sets of parameters $\{u_i\}$ and $\{v_i\}$.

~~\\
\centerline{{\bf 9. Equivalence of the transfer matrix eigenvalues}}
~~\\

The above calculations show that there are three different forms for the 
Bethe ansatz solution of this model. Here we will argue that 
these forms are in fact equivalent. The method that we will use is 
inspired by the paper \cite{pf97} which hints at the argument we 
will give without providing many details. Our goal here is to make the 
argument more transparent.  

We begin with the following $R$-matrix \cite{bdgz,dglz} which
 is associated with 
the quantum superalgebra $U_q(sl(1|1))$. The solution we present below 
is more general than that of \cite{bdgz,dglz} in that we have included a gauge
parameter $l$ which is necessary for our argument to succeed. The 
$R$-matrix is 
\be r(v,l)^{\a\b}=\pmatrix{[v-(\a+\b)/2-1]&0&0&0 \cr 0&l[v+(\a-\b)/2]& q^{v}
\sqrt{[\a+1][\b+1]}&0 \cr
0&-q^{-v}\sqrt{[\a+1][\b+1]}& l^{-1}[v+(\b-\a)/2] &0 \cr 0&0&0& 
[v+(\a+\b)/2+1] }
\label{6v} \ee
which satisfies the coloured Yang-Baxter equation 
$$r^{\a\b}_{12}(u-v,l)r^{\a\g}_{13}(u,l)r^{\b\g}_{23}(v,l)
=r^{\b\g}_{23}(v,l)r^{\a\g}_{13}(u,l)r^{\a\b}_{12}(u-v,l).$$
We emphasize again that the tensor products in the above equation are to
be evaluated according to the rule (\ref{rule}). In (\ref{6v}) we have
adopted the convention that index 1 is odd and 2 is even.

Using this solution we can construct the transfer matrix

$$\tau(v,\a,\b,u_i,l,x_1,x_2)={\rm str}_0\l(X_0
r^{\a0}_{01}(v-u_1,l)...r^{\a0}_{0P}(v-u_P,l)
r^{\a\b}_{0(P+1)}(v,l)...r^{\a\b}_{0(L+P)}(v,l)\r) $$ 
where $X={\rm diag}(x_1,x_2),\,x\in \mathbb C$. These matrices form a 
commuting family with the property
$$[\tau(v,\a,\b,u_i,l,x_1,x_2),\,\tau(u,\a,\g,u_i,l,x_1,x_2)]=0.$$ 

For the diagonalization of the above transfer matrix by the Bethe ansatz
there are two available approaches. We may begin with a reference state
given by 
$$(v^2)^{\otimes (L+P)}$$
in which case the eigenvalues read
\bea  \lambda^-(v,\a,\b,u_i,w_j,l,x_1,x_2)=\l(x_2[v+(\a+\b)/2+1]^L
\prod_{i=1}^P\r.&&[v-u_i+\a/2+1] \nn \\
 \l. -x_1l^L[v+(\a-\b)/2]^L
\prod_{i=1}^Pl[v-u_i+\a/2]\r) 
 \prod_{j=1}^M &&\frac{[w_j-v+\a/2+1/2]}     
{l[w_j-v-\a/2-1/2]} \label{eeig1} \eea   
such that the parameters $w_j$ satisfy the Bethe ansatz equations 
\be x_2\l(\frac{[w_j+\b/2+1/2]}{[w_j-\b/2-1/2]}\r)^L=x_1l^L\prod_{i=1}^P
\frac{l[u_i-w_j+1/2]}{[u_i-w_j-1/2]}.\label{ebae1} \ee 

We can also use
$$(v^1)^{\otimes (L+P)}$$ 
for the pseudo-vacuum which yields the eigenvalue expression
\bea \lambda^+(v,\a,\b,u_i,\w_j,l,x_1,x_2)=-
\l(x_1[v-(\a+\b)/2-1]^L \prod_{i=1}^N\r.&&[v-u_i-\a/2-1] \nn \\
 \l.-x_2l^{-L}[v+(\b-\a)/2]^L
\prod_{i=1}^Pl^{-1}[v-u_i-\a/2]\r) 
\prod_{j=1}^{L+P-M}&&\frac{l[\w_j-v-\a/2-1/2]}
 {[\w_j-v+\a/2+1/2]} \label{eeig2}\eea  
such that the parameters $\w_j$ satisfy the Bethe ansatz equations
\be 
x_2\l(\frac{[\w_j+\b/2+1/2]}{[\w_j-\b/2-1/2]}\r)^L=x_1l^L
\prod_{i=1}^{P}
\frac{l[u_i-\w_j+1/2]}{[u_i-\w_j-1/2]} \label{ebae2}\ee



We conjecture that for each set of $M$ parameters $\{w_j\}$ satisfying 
(\ref{ebae1}), there is a set of (L+P-M) parameters $\{\w_i\}$ satisfying
(\ref{ebae2}) which render (\ref{eeig1}) and (\ref{eeig2}) equal. 
In the rational limit $q\rightarrow 1$ this argument can be made 
rigorous since in this instance there is an underlying $sl(1|1)$ 
invariance for the transfer matrix. As mentioned before, one can 
prove a highest and lowest weight theorem respectively for the 
two sets of Bethe states and then use a combinatorial argument to 
claim that the multiplets generated by the Bethe states give the 
full space of states in each case, so there 
must be a one-to-one correspondence. 
Although this proof will fail for generic
values of $q$, we still believe that the result holds true.

Applying the conjecture, it then follows that  
\bea &&\Lambda_3(v,u_i,w_j) \nn \\
&=&\frac{1}{[v+\a]^L}
\prod_{j=1}^{L-n_-}\frac{1}{[v-u_i-\a/2-1/2]}
\nn \\
&&\times\l\{
-\lambda^-\l(v,\a-1,\a-1,u_i,w_j,p_2,\l(\frac{p_1}{p_2}\r)^{(L-n_-)},
\l(\frac{p_1p_3}{p_4}\r)^{(L-n_-)}\r)
\r. \nn \\
&&\l.
+\frac{[v]^L}{[v-\a-1]^L}
\lambda^-\l(v-1/2,\a,\a-1,u_i,w_j,\frac{p_2p_4}{p_3},p_1^{-n_-}
\l(\frac{p_2p_4}{p_3}\r)^{(n_--L)},\l(\frac{p_1p_3}{p_4}\r)^{-n_-}\r)\r\}\nn \\
&=&\frac{1}{[v+\a]^L}\prod_{j=1}^{L-n_-}\frac{1}{[v-u_i-\a/2-1/2]}
\nn \\
&&\times \l\{-\lambda^+\l(v,\a-1,\a-1,u_i,\w_j,p_2,\l(\frac{p_1}{p_2}\r)^{(L-n_-)},
\l(\frac{p_1p_3}{p_4}\r)^{(L-n_-)},\l(\frac{p_1p_3}{p_4}\r)^{-n_-}\r)
\r. \nn \\
&&\l.
+\frac{[v]^L}{[v-\a-1]^L}
\lambda^+\l(v-1/2,\a,\a-1,u_i,\w_j,\frac{p_2p_4}{p_3},p_1^{-n_-}
\l(\frac{p_2p_4}{p_3}\r)^{(n_--L)},\l(\frac{p_1p_3}{p_4}\r)^{-n_-}\r)\r\}\nn
\\
&=&\Lambda_1(v,\w_j,u_i)    \\   \eea
Similarly, we find 
\bea &&\oL_3(v,u_i,w_j) \\
&=&\frac{1}{[v+\a]^L}
\prod_{j=1}^{L-n_+}\frac{1}{[v-u_i-\a/2-1/2]}
\nn \\
&&\times\l\{
-\lambda^-\l(v,\a-1,\a-1,u_i,w_j,p_1,\l(\frac{p_2}{p_1}\r)^{(L-n_+)},
\l(\frac{p_2p_4}{p_3}\r)^{(L-n_+)}\r)
\r. \nn \\
&&\l.
+\frac{[v]^L}{[v-\a-1]^L}
\lambda^-\l(v-1/2,\a,\a-1,u_i,w_j,\frac{p_1p_3}{p_4},p_2^{-n_+}
\l(\frac{p_1p_3}{p_4}\r)^{(n_+-L)},\l(\frac{p_2p_4}{p_3}\r)^{-n_+}\r)\r\}\nn
\\
&=&\frac{1}{[v+\a]^L}\prod_{j=1}^{L-n_+}\frac{1}{[v-u_i-\a/2-1/2]}
\nn \\
&&\times
\l\{-\lambda^+\l(v,\a-1,\a-1,u_i,\w_j,p_1,\l(\frac{p_2}{p_1}\r)^{(L-n_+
)},
\l(\frac{p_2p_4}{p_3}\r)^{(L-n_+)},\l(\frac{p_2p_4}{p_3}\r)^{-n_+}\r)
\r. \nn \\
&&\l.
+\frac{[v]^L}{[v-\a-1]^L}
\lambda^+\l(v-1/2,\a,\a-1,u_i,\w_j,\frac{p_1p_3}{p_4},p_2^{-n_+}
\l(\frac{p_1p_3}{p_4}\r)^{(n_+-L)},\l(\frac{p_2p_4}{p_3}\r)^{-n_+}\r)\r\}\nn
\\
&=&\oL_1(v,\w_j,u_i)    \\   \eea
although the result is  immediate through use of (\ref{inv}).
In a similar fashion it can be shown that 
$$ \La_3(v,u_i,w_j)=\La_2(v,\u_i,w_j),~~~\oL_3(v,u_i,w_j)=
\oL_2(v,\u_i,w_j)
$$
though the judicious mathematician would appeal to (\ref{inv1}) 
to obtain the result.

~~\\
\centerline{{\bf 10. Quantum transfer matrix eigenvalues}}
~~\\

The Bethe ansatz approach for the diagonalization of the quantum
transfer matrix proceeds in much the same way as the previous
discussion. We return to the algebraic relations 
(\ref{yba1},\ref{yba2},\ref{yba3}). A
representation of the algebras $Y$ and $Z$ are obtained through 
\bea \pi\l(Y(v)\r)&=&\left(L^{\mathrm{st}_{\L}}_{\L0}(\chi-v)
L_{0(\L-1)}(\chi+v)....L^{\mathrm{st}_2}_{20}(\chi-v)L_{01}(\chi+v)\right)\nn \\
\pi\l(Z(v)\r)&=&\left(R^{\mathrm{st}_{\L}}_{\L0}(\chi-v)
R_{0(\L-1)}(\chi+v)....R^{\mathrm{st}_2}_{20}(\chi-v)R_{01}(\chi+v)\right)
\nn  \eea 
where the label 0 refers to the auxiliary space (three dimensional for
$L(v)$ and four dimensional for $R(v)$) and
the natural numbers label quantum spaces. The quantum transfer matrix is
given by 
$$Q(v)= \sum_{i=1}^4(-1)^{(i)+(i)(k)}\pi\left(Z^i_i(v)\right)^k_l$$
in complete analogy with (\ref{grad}). The principle difference is in the
choice of the reference state. In order to adopt the ansatz (\ref{ansatz}) we
need to find a reference state which satisfies the conditions (\ref{cond}).
This is achieved with the choice 
$$\Phi^0=(v^1\ot v^4)^{\ot \L/2}$$  
which is an eigenstate of the quantum transfer matrix with eigenvalue
\bea &&\l(\frac{[\chi+v-\a][\chi-v][\chi-v-1]}{p_1p_2[\chi+v+ 
\a][\chi-v+\a][\chi-v-\a-1]}\r)^{\L/2}-\l(\frac{p_3[\chi+v] [\chi-v]}
{p_1p_2p_4[\chi+v+\a][\chi-v-\a-1]}\r)^{\L/2} \nn \\
&&
-\l(\frac{p_4[\chi+v][\chi-v]}{p_1p_2p_3[\chi+v+\a][\chi-v-\a-1]}\r)^{\L/2}
+\l(\frac{[\chi+v][\chi+v-1][\chi-v+\a+1]}
{p_1p_2[\chi+v+\a][\chi+v-\a-1][\chi-v-\a-1]}\r)^{\L/2}.  \nn \eea 
At this point we can follow the Bethe ansatz procedure exactly as
before. This yields the eigenvalue expression for the quantum transfer
matrix 
\bea && \La(v)= \nn \\
&&(p_1p_2^{-1}p_3^2p_4^{-2})^{\M} 
\l(\frac{[\chi+v][\chi+v-1][\chi-v+\a+1]}{p_1p_2[\chi+v+ 
\a][\chi+v-\a-1][\chi-v-\a-1]}\r)^{\L/2}\prod_{i=1}^{\N}
\frac{p_2p_4[u_i-v-\a/2]}{p_3[u_i-v+\a/2+1]} \nn \\
&&\l(\frac{p_3[\chi+v] [\chi-v]}
{p_1p_2p_4[\chi+v+\a][\chi-v-\a-1]}\r)^{\L/2}
\prod_{i=1}^{\N}\l(\frac{p_2p_4[u_i-v-\a/2]}
{p_3[u_i-v+\a/2+1]}\r)\prod_{j=1}^{\M}\l(\frac{p_1p_3}{p_2p_4}
\frac{[w_j-v+\a/2+3/2]}{[w_j-v+\a/2+1/2]}\r) \nn \\
&&-\l(\frac{p_4[\chi+v][\chi-v]}
{p_1p_2p_3[\chi+v+\a][\chi-v-\a-1]}\r)^{\L/2}
\prod_{i=1}^{\N}\l(\frac{p_2[u_i-v-\a/2]}{
[u_i-v+\a/2]}\r) \prod_{j=1}^{\M}\l(\frac{p_1p_3}{p_2p_4}
\frac{[w_j-v+\a/2-1/2]}{[w_j-v+\a/2+1/2]}\r)
\nn \\
&&+(p_1p_2^{-1})^{-\M}\l(\frac{[\chi+v-\a][\chi-v][\chi-v-1]}
{p_1p_2[\chi+v+\a][\chi-v+\a][\chi-v-\a-1]}\r)^{\L/2}
\prod_{i=1}^{\N}\frac{p_2[u_i-v-\a/2]}{[u_i-v+\a/2]}
\nn \eea 
subject to the Bethe ansatz equations 
\bea 
\left(\frac{p_3[\chi+u_i-\a/2][\chi-u_i-\a/2-1]}{p_4[\chi+u_i+\a/2]
[\chi-u_i+\a/2]}\right)^{\L/2}&=&\prod_{j=1}^{\M}\frac{p_3[w_j-u_i-1/2]}
{p_4[w_j-u_i+1/2]} \nn \\
\left(\frac{p_4}{p_3}\right)^{\L}\prod_{i=1}^{\N}\frac{p_3[u_i-w_j+1/2]}
{p_4[u_i-w_j-1/2]}&=&-\prod_{k=1}^{\M}\frac{[w_j-w_k-1]}{[w_j-w_k+1]}. \nn 
\eea
Another expression is
obtained through (\ref{inv}). It is curious that the gauge parameters
$p_1,\,p_2$ do not appear in these Bethe ansatz equations. 
As explained earlier, equivalent forms, which
we will not make explicit here, can be obtained through the use of
the equivalence of the expressions (\ref{eeig1},\ref{eeig2}). 

~~\\
\centerline{{\bf Acknowledgements}}
~~\\

JL is supported by an Australian Research Fellowship awarded through the
Australian Research Council.
AF thanks CNPq-Conselho Nacional de Desenvolvimento Cient\'{\i}fico e
Tecnol\'ogico for financial support.
We also thank the Max-Planck-Institut f\"ur Physik Komplexer 
Systeme, Dresden, 
where this work was initiated at the meeting Cooperative Phenomena
in Statistical Physics: Theory and Applications.



\begin{thebibliography}{99}

\bibitem{j} M. Jimbo, Lett. Math. Phys. {\bf 10} (1985) 63. 
\bibitem{d1} V.G. Drinfeld, {\it Quantum Groups} in Proceedings of 
the International Congress of Mathematicians, ed. A.M. Gleason 
(American Mathematical Society, Providence, 1986) 798. 
\bibitem{bgz} A.J. Bracken, M.D. Gould and R.B. Zhang,
Mod. Phys. Lett. A {\bf 5} (1990) 831.
\bibitem{manfred} M. Scheunert, Lett. Math. Phys. {\bf 24} (1992) 173;
J. Math. Phys. {\bf 34} (1993) 3780.
\bibitem{y} H. Yamane, Publ. Proc. Japan Acad. {\bf 70} (1991) 31;
Publ. RIMS Kyoto Univ. {\bf 30} (1994) 15.
\bibitem{kt} S.M. Khoroshkin and V.N. Tolstoy, Commun. Math. Phys. {\bf
141} (1991) 599.
\bibitem{fst} L.D. Faddeev, E.K. Sklyanin and L.A. Takhtajan, Theor.
Math. Phys. {\bf 40} (1979) 194; \\
P.P. Kulish and E.K. Sklyanin, Lect. Notes in Phys. {\bf 151} (1982) 61.
\bibitem{d2} V.G. Drinfeld, Leningrad Math. J. {\bf 1} (1990) 1419.
\bibitem{re} N.Y. Reshetikhin, Lett. Math. Phys. {\bf 20} (1990) 331.  
\bibitem{flr} A. Foerster, J. Links and I. Roditi, J. Phys. A: 
Math. Gen. {\bf 31} (1998) 687.
\bibitem{ek} R.A. Engeldinger and A. Kempf, J. Math. Phys. {\bf 35} 
(1994) 1931.
\bibitem{fv} F.H.L. Essler and V.E. Korepin, Phys. Rev. B {\bf 46}
(1992) 9147.
\bibitem{fk} A. Foerster and M. Karowski, Phys. Rev. B {\bf 46} (1992)
9234; Nucl. Phys. B {\bf 396} (1993) 611.
\bibitem{eks} F.H.L. Essler, V.E. Korepin and K. Schoutens, Int. 
J. Mod. Phys. B {\bf 8} (1994) 3205.
\bibitem{pf96} M.P. Pfannm\"uller and H. Frahm, Nucl. Phys. B {\bf
479} (1996) 575.
\bibitem{pf97} M.P. Pfannm\"uller and H. Frahm, J. Phys. A: Math. 
Gen. {\bf30} (1997) L543. 
\bibitem{lf99} J. Links and A. Foerster, J. Phys. A: Math. Gen. 
{\bf 32} (1999) 147.
\bibitem{flt} A. Foerster, J. Links and A.P. Tonel, Nucl. Phys. B
{\bf 552} (1999) 707.
\bibitem{fss} L. Frappat, A. Sciarrino and P. Sorba, Commun. Math. Phys.
{\bf 121} (1989) 457.
\bibitem{lai} C.K. Lai, J. Math. Phys. {\bf 15} (1974) 1675.
\bibitem{s} B. Sutherland, Phys. Rev. B {\bf 12} (1975) 3795.
\bibitem{ar} J. Abad and M. R\'{\i}os, Phys. Rev. B {\bf 53} (1996) 
14000; J. Phys. A: Math. Gen. {\bf 30} (1997) 5887. 
\bibitem{bkz} R.Z. Bariev, A. Kl\"umper and J. Zittartz, Europhys.
Lett.  {\bf 32} (1995) 85.
\bibitem{ba} R.Z. Bariev, J. Phys. A: Math. Gen. {\bf 24} (1991) L549 
\bibitem{bglz} A.J. Bracken, M.D. Gould, J.R. Links and Y.-Z. Zhang, 
Phys. Rev. Lett. {\bf 74} (1995) 2768. 
\bibitem{ghlz} M.D. Gould, K.E. Hibberd, J.R. Links and Y.-Z. Zhang,
Phys. Lett. A {\bf 212} (1996) 156.
\bibitem{bdgz} A.J. Bracken, G.W. Delius, M.D. Gould and Y.-Z. Zhang,
J. Phys. A: Math. Gen. {\bf 27} (1994) 6551.
\bibitem{dglz} G.W. Delius, M.D. Gould, J.R. Links and Y.-Z. Zhang,
Int. J. Mod. Phys. A {\bf 10} (1995) 3259.
\bibitem{m} Z. Maassarani, J. Phys A: Math. Gen {\bf 28} (1995) 1305. 
\bibitem{a} D. Arnaudon, J. High Energy Phys. {\bf 12} (1997) 6. 
\bibitem{b} H.M. Babujian, Nucl. Phys. B {\bf 215} (1983) 317.
\bibitem{bt} H.M. Babujian and A.M. Tsvelick, Nucl. Phys. B {\bf 265}
(1986) 24. 
\bibitem{lfk} J. Links, A. Foerster and M. Karowski, J. Math. Phys. 
{\bf 40} (1999) 726.
\bibitem{g99} J. Gruneberg, Commun. Math. Phys. {\bf 206} (1999) 383.
\bibitem{g00} J. Gruneberg, Nucl. Phys. B {\bf 568} (2000) 594.
\bibitem{rm} P.B. Ramos and M.J. Martins, Nucl. Phys. B 
{\bf 479} (1996) 678.
\bibitem{t} Z. Tsuboi, J. Phys. A: Math. Gen. {\bf 31} (1998) 5485. 
\bibitem{k} A. Kl\"umper, Ann. Physik {\bf 1} (1992) 540; Z. Phys. 
B {\bf 91} (1993) 507. 
\bibitem{jks} G. J\"uttner, A. Kl\"umper and J. Suzuki, J. Phys. A: 
Math. Gen. {\bf 30} (1997) 1881.
\bibitem{kwz} A. Kl\"umper, T. Wehner and J. Zittartz, J. Phys. A: 
Math. Gen. {\bf 30} (1997) 1897.
\bibitem{lf97} J. Links and A. Foerster, J. Phys. A: Math. Gen. 
{\bf 30} (1997) 2483. 
\bibitem{jc} A.D. Jacobs and J.F. Cornwell, J. Math. Phys. {\bf 38}
(1997) 5383. 
\bibitem{l} J. Links, J. Phys. A: Math. Gen. {\bf 32} (1999) L315.
\bibitem{schultz} C.L. Schultz, Physica A {\bf 122} (1983) 71.
%
%
%
\end{thebibliography}
\end{document}